\documentclass[a4paper,fleqn,usenatbib]{mnras}

\usepackage[T1]{fontenc}
\usepackage{ae,aecompl}

\usepackage{graphicx}	
\usepackage{amsmath}	
\usepackage{amssymb}	
\usepackage{xcolor}
\usepackage{rotating}  
\usepackage{pdflscape}

\newcommand{\PSRnumNCB}{60 } 
\newcommand{\PSRnumADC}{40 } 
\newcommand{\PSRnumHTRUTOTAL}{100 } 
\newcommand{\PSRnumADCconfbin}{8 } 
\newcommand{\PSRnumADCunsolved}{7 } 
\newcommand{\PSRnumADCsolved}{33 } 
\newcommand{\PSRnumNCBsolved}{31 } 
\newcommand{\PSRnumTOTALsolved}{64 } 
\newcommand{\PSRnumADCsolvedminusbinaries}{28 } 
\newcommand{\PSRnumADCdfbpks}{27 } 
\newcommand{\PSRnumADCdfbpksminusmsp}{22 } 
\newcommand{\PSRnumADCunsummed}{13 } 
\newcommand{\PSRnumADCPMPStotal}{22 } 
\newcommand{\PSRnumADCPMPSovereight}{13 } 
\newcommand{\PSRnumADCPMPSundereight}{9 } 
\newcommand{\PSRnumKNOWNsurveyPPDOT}{817 } 
\newcommand{\PSRnumKNOWNsurveyDISTANCE}{835 } 
\newcommand{\PSRnumKNOWNsurveyAGE}{756 } 
\newcommand{\PSRnumHTRUAGE}{56 } 
\newcommand{\PSRnumHTRUFLUX}{81 } 
\newcommand{\PSRnumKNOWNsurveyFLUX}{748 } 
\newcommand{\PSRindividualREDETECTIONS}{755 }
\newcommand{\PSRuniqueREDETECTIONS}{390 }

\title[HTRU$-$XVI. 40 new pulsars in the Galactic plane.]{The High Time Resolution Universe Pulsar Survey - XVI. Discovery and timing of \PSRnumADC pulsars from the southern Galactic plane.}

\author[A.~D.~Cameron et~al.]{
A.~D.~Cameron$^{1,2}$\thanks{E-mail: andrew.cameron@csiro.au},
D.~J.~Champion$^{1}$,
M.~Bailes$^{3,4}$,
V.~Balakrishnan$^{1}$,
\newauthor
E.~D.~Barr$^{1,4}$,
C.~G.~Bassa$^{5}$,
S.~Bates$^{6}$,
S.~Bhandari$^{2}$,
N.~D.~R.~Bhat$^{7}$,
M.~Burgay$^{8}$,
\newauthor
S.~Burke-Spolaor$^{9}$,
C.~M.~L.~Flynn$^{3}$,
A.~Jameson$^{3,4}$,
S.~Johnston$^{2}$,
M.~J.~Keith$^{6}$,
\newauthor
M.~Kramer$^{1,6}$,
L.~Levin$^{6}$,
A.~G.~Lyne$^{6}$,
C.~Ng$^{10}$,
E.~Petroff$^{5}$,
A.~Possenti$^{8,11}$,
\newauthor
D.~A.~Smith$^{12}$,
B.~W.~Stappers$^{6}$,
W.~van~Straten$^{13}$,
C.~Tiburzi$^{1,14}$,
J.~Wu$^{1}$
\\
$^{1}$Max-Planck Institut f{\"u}r Radioastronomie, Auf dem H{\"u}gel 69, D-53121 Bonn, Germany\\
$^{2}$CSIRO Astronomy \& Space Science, Australia Telescope National Facility, P.O. Box 76, Epping, NSW 1710, Australia.\\
$^{3}$Centre for Astrophysics and Supercomputing, Swinburne University of Technology, Mail H39, PO Box 218, VIC 3122, Australia.\\
$^{4}$ARC Center of Excellence for Gravitational Wave Discovery (OzGrav), Swinburne University of Technology, Mail H11, PO Box 218, VIC 3122, Australia.\\
$^{5}$ ASTRON, the Netherlands Institute for Radio Astronomy, Postbus 2, NL-7990 AA Dwingeloo, the Netherlands.\\
$^{6}$Jodrell Bank Center for Astrophysics, University of Manchester, Alan Turing Building, Oxford Road, Manchester M13 9PL, United Kingdom.\\
$^{7}$International Centre for Radio Astronomy Research, Curtin University, Bentley, WA 6102, Australia.\\
$^{8}$INAF - Osservatorio Astronomico di Cagliari, Via della Scienza 5, I-09047 Selargius (CA), Italy.\\
$^{9}$Center for Gravitational Waves and Cosmology, West Virginia University, Chestnut Ridge Research Building, Morgantown, WV 26505, USA.\\
$^{10}$Department of Physics and Astronomy, University of British Columbia, 6224 Agricultural Road, Vancouver, BC V6T 1Z1, Canada.\\
$^{11}$Universit\'a di Cagliari, Dept of Physics, S.P. Monserrato-Sestu Km 0,700 - 09042 Monserrato, Italy \\
$^{12}$Centre d'\'{E}tudes Nucl\'{e}aires de Bordeaux Gradignan, IN2P3/CNRS, Universit\'{e} Bordeaux 1, BP120, F-33175 Gradignan Cedex, France \\
$^{13}$Institute for Radio Astronomy \& Space Research, Auckland University of Technology, Private Bag 92006, Auckland 1142, New Zealand.\\
$^{14}$Fakult{\"a}t f{\"u}r Physik, Universit{\"a}t Bielefeld, Postfach 100131, D-33501 Bielefeld, Germany.\\
}

\date{Accepted XXX. Received YYY; in original form ZZZ}

\pubyear{2020}

\begin{document}
\label{firstpage}
\pagerange{\pageref{firstpage}--\pageref{lastpage}}
\maketitle

\begin{abstract}
We present the results of processing an additional 44\,\% of the High Time Resolution Universe South Low Latitude (HTRU-S LowLat) pulsar survey, the most sensitive blind pulsar survey of the southern Galactic plane to date. Our partially-coherent segmented acceleration search pipeline is designed to enable the discovery of pulsars in short, highly-accelerated orbits, while our 72-min integration lengths will allow us to discover pulsars at the lower end of the pulsar luminosity distribution. We report the discovery of \PSRnumADC pulsars, including three millisecond pulsar-white dwarf binary systems (PSRs~J1537$-$5312, J1547$-$5709 and J1618$-$4624), a black-widow binary system (PSR~J1745$-$23) and a candidate black-widow binary system (PSR~J1727$-$2951), a glitching pulsar (PSR~J1706$-$4434), an eclipsing binary pulsar with a 1.5-yr orbital period (PSR~J1653$-$45), and a pair of long spin-period binary pulsars which display either nulling or intermittent behaviour (PSRs~J1812$-$15 and J1831$-$04). We show that the total population of 100 pulsars discovered in the HTRU-S LowLat survey to date represents both an older and lower-luminosity population, and indicates that we have yet to reach the bottom of the luminosity distribution function. We present evaluations of the performance of our search technique and of the overall yield of the survey, considering the 94\,\% of the survey which we have processed to date. We show that our pulsar yield falls below earlier predictions by approximately 25\,\% (especially in the case of millisecond pulsars), and discuss explanations for this discrepancy as well as future adaptations in RFI mitigation and searching techniques which may address these shortfalls.
\end{abstract}

\begin{keywords}
surveys -- stars: neutron -- pulsars: general
\end{keywords}



\section{Introduction}\label{sec: Introduction}

The ongoing search for new pulsars remains a core goal of current efforts in pulsar astronomy. Although the ATNF Pulsar Catalogue\footnote{http://www.atnf.csiro.au/people/pulsar/psrcat/} \citep[\textsc{psrcat},][]{mhth05} already lists approximately 2800 pulsars, the body of which have allowed for enormous progress in our understanding of multiple areas of physics, it is largely by the discovery of new pulsars that our understanding is extended. New discoveries help complete our knowledge of the underlying pulsar population, and the more unusual and exotic of these new pulsars present new physical challenges to be explored and explained.

We have therefore undertaken the High Time Resolution Universe South Low Latitude (HTRU-S LowLat) pulsar survey, a long-integration, high-time resolution blind pulsar survey of the southern Galactic plane region taken with the 21-cm multibeam receiver \citep[MB20;][]{swb+96} of the Parkes 64-m Radio Telescope. The survey is one component of the HTRU-South all-sky pulsar survey \citep[HTRU-S;][]{kjvs10}, which fully covers the region of sky below a declination of $\delta<+10^\circ$. The three components of this survey are divided into regions of Galactic longitude $l$ and latitude $b$, giving the High, Mid and Low latitude survey regions, with the design of each survey tailored towards specific scientific objectives. HTRU-S is complemented by a corresponding northern survey, HTRU-North \citep[HTRU-N;][]{bck+13}, undertaken with the Effelsberg 100-m Radio Telescope.

While the general goal of any pulsar survey can largely be characterised as the discovery of new pulsars, the primary goal of HTRU-S LowLat has been the discovery of relativistic binary pulsars. These binary systems can serve as excellent laboratories for developing tests of gravitational theories such as general relativity (GR), with the Double Pulsar PSR~J0737$-$3039 \citep{bdp+03,lbk+04,ksm+06} currently standing as the leading example of such a gravitational laboratory. The region of sky surveyed by HTRU-S LowLat (between Galactic longitudes of $-80^\circ<l<30^\circ$ and Galactic latitudes $\left|b\right|<3.5^\circ$) comprises the densest portion of the Galactic plane, and is expected to contain the highest proportion of these systems \citep{bkb02}. Compact binary systems with short orbital periods ($P_\text{b}<12\,\text{h}$) are of particular interest, as these systems are likely to display the most significant relativistic effects, allowing for new and improved tests and limits well beyond those available from current binary pulsars.

An important secondary goal of the survey involves the discovery of low-luminosity pulsars too weak to have been detected by earlier pulsar surveys. This is made possible by the 72-min observations employed by HTRU-S LowLat. Significant effort has been spent attempting to model the luminosity distribution function of pulsars both within and outside the Galactic plane \citep[see e.g.][]{fk06, lbb+13, blrs14, gmv+14,cbo18,ghf+18}, as well as to link this distribution to other aspects of pulsar phenomenology (e.g. spin periods and spin-period derivatives). However, such efforts are intrinsically constrained by those pulsars which are currently known and available for study. Further characterising the population statistics of low-luminosity pulsars will allow for a greater understanding of the pulsar population as a whole, and will be vital in the planning of future pulsars surveys to be undertaken with next generation radio telescopes such as MeerKAT\footnote{http://www.ska.ac.za/science-engineering/meerkat/}, the Square Kilometre Array\footnote{http://skatelescope.org/} (SKA) and the Five-hundred-meter Aperture Spherical Telescope\footnote{http://fast.bao.ac.cn/en/} (FAST), which are likely to probe even deeper into this regime. Finally, the long observation times are also favorable for capturing various transient radio phenomena, including Rotating Radio Transients \citep[RRATs; see e.g.][]{mll+06,kle+10}, both nulling and intermittent pulsars, and potentially Fast Radio Bursts \citep[FRBs; see e.g.][]{lbm+07, tsb+13, cpk+16}.

The HTRU-S LowLat survey has already made a significant scientific contribution to each of these scientific goals, and this paper marks its third major discovery publication. \cite{ncb15} reported the discovery of \PSRnumNCB pulsars in the first 50\,\% of the survey to be processed, along with the specifications of the `partially-coherent segmented acceleration search' pipeline used to analyse the data (further described in Section\,\ref{sec: methodology}). The reported population of pulsars were of lower luminosity than the background population from the survey region and, in the case of the un-recycled pulsars, appeared to represent a typically older population. \cite{ncb15} also reported four binary pulsars (PSRs~J1101$-$6424, J1244$-$6359, J1755$-$25 and J1759$-$24) and two nulling pulsars (PSRs~J1227$-$63 and J1349$-$63), demonstrating HTRU-S LowLat's ability to discover both binary systems and transient phenomena. Meanwhile, continued processing of the remaining portion of the survey with the same search pipeline resulted in the discovery of PSR~J1757$-$1854 as reported in \cite{cck+18}, the most accelerated binary pulsar currently known and the only relativistic binary pulsar to have been discovered in the entire HTRU-S survey thus far.

This paper presents new results derived from the processing of an additional 44\,\% of the HTRU-S LowLat survey through the `partially-coherent segmented acceleration search' as described in \cite{ncb15}, including the discovery of a further \PSRnumADC pulsars\footnote{This total includes the previously-published PSR~J1757$-$1854, which is included here as it was discovered as part of survey processing presented in this paper.\label{ftnt: J1757-footnote}}. This brings the total fraction of the survey processed through this pipeline to 94\,\%. Section~\ref{sec: methodology} summarises the search strategy used in the processing of the survey data, and the current status of the survey processing. Section~\ref{sec: redetections} discusses the redetections of known pulsars within the processed 44\,\% of the survey data. Section~\ref{sec: new discoveries} presents the details of the newly-discovered pulsars, including timing solutions where available. Section~\ref{sec: pulsars of interest} then follows-up with an in-depth discussion of a selection of pulsars of interest. Section~\ref{sec: pop comparison} provides an analysis of the survey discoveries within the context of the larger pulsar population, with an evaluation of the survey yield presented in Section~\ref{sec: survey yield}. Additional discussion and conclusions follow in Section~\ref{sec: conclusions}.

\section{Methodology}\label{sec: methodology}

The `partially-coherent segmented acceleration search' pipeline used to process the 44\,\% of the HTRU-S LowLat survey analysed for this paper (as well as the subsequent pulsar candidate identification and confirmation procedures) are the same as those described by \cite{ncb15}. For completeness, we present here a brief summary of this technique.

In order to allow for the detection of binary pulsars, our pipeline employs the `time-domain resampling' acceleration search technique \citep[see e.g.][]{mk84,jk91}. This technique assumes that the orbital motion of a pulsar can be modeled by a constant acceleration $a$ over the span of a given observation such that the line-of-sight velocity can be expressed by $V\left(t\right)=at$. If this assumption remains true (i.e. that the `jerk' or rate of change of acceleration $\dot{a}\simeq0\,\text{m}\,\text{s}^{-3}$), then each dedispersed time series (up to a maximum dispersion measure (DM) of $3000\,\text{cm}^{-3}\,\text{pc}$) can be quadratically resampled so as to remove the effect of the orbital motion over the course of the observation. This technique works best (in the case of circular orbits) when the parameter
\begin{equation}\label{eqn: rorb}
r_\text{orb} = \frac{t_\text{int}}{P_\text{b}} \lesssim 0.1,
\end{equation}
where $t_\text{int}$ is the integration time of the observation and $P_\text{b}$ is the orbital period of the pulsar. This value of $r_\text{orb}$ is based on an in-depth investigation detailed in Appendix~A1 of \cite{ncb15}.

The integration time of each observation is $t_\text{int}=4300\,\text{s}$, which by Equation~\ref{eqn: rorb} implies a sensitivity to $P_\text{b} \gtrsim 12\,\text{h}$. In order to optimise the sensitivity of each observation to shorter orbital periods, we adopt a segmented search strategy. Each observation is broken into full-length ($s=1$), half-length ($s=2$), quarter-length ($s=4$) and eighth-length ($s=8$) segments, resulting in 15 segments in total. Each group of segments spans the entire full-length observation without overlap. This segmenting provides sensitivity to progressively shorter orbital periods, with the $t_\text{int}$ of each segment and its optimal $P_\text{b}$ range described in Table~\ref{tab: segment parameters}. However, with each additional halving of the observation length $t_\text{int}$, the flux-density sensitivity of the segments correspondingly lowers by a factor of $\sqrt{2}$. Our strategy aims to strike an ideal balance between these two considerations, optimising our ability to detect binary systems with small values of $P_\text{b}$ while retaining as much of the observation length and therefore as much sensitivity as possible. Each of the 15 segments of a given observation beam is searched coherently through an acceleration and Fourier search, but is processed independently to each of the other segments, hence rendering the pipeline only `partially-coherent'.

\begin{table}
\begin{center}
\caption{A summary of the 15 individual segments searched as part of the partially-coherent segmented acceleration search, including the number of segments ($s$) in each group, the minimum orbital period ($P_\text{b}$) to which each is sensitive and their acceleration search ranges ($\left|a_\text{min}\right|$ to $\left|a_\text{max}\right|$).}\label{tab: segment parameters}
\begin{tabular}{cccccc}
\hline
Segment & $s$ & $t_\text{int}$ & Min. $P_\text{b}$ & $\left|a_\text{min}\right|$ & $\left|a_\text{max}\right|$ \\
 & & (s) & (h) & ($\text{m}\,\text{s}^{-2}$) & ($\text{m}\,\text{s}^{-2}$) \\
\hline
Full-length & 1 & 4300 & 12 & $^{\phantom{a}}$0$^{\phantom{a}}$ & 1 \\
Half-length & 2 & 2150 & 6 & $^{\phantom{a}}$0$^{\phantom{a}}$ & 200 \\
Quarter-length & 4 & 1075 & 3 & $^{\phantom{a}}$200$^{\text{a}}$ & 500 \\
Eighth-length & 8 & 537 & 1.5 & $^{\phantom{a}}$0$^{\phantom{a}}$ & 1200 \\
\hline
\multicolumn{6}{l}{$^{\text{a}}$A minority of the processed data was also searched} \\
\multicolumn{6}{l}{within $\left|a\right| < 200\,\text{m\,s}^{-2}$ in the s=4 segment.} 
\end{tabular}
\end{center}
\end{table}

Acceleration search ranges for the $s=2,4,8$ segments are chosen by adopting a hypothetical neutron star-black hole (NS-BH) binary scenario as a limiting case. Assuming a circular orbit with a $1.4\,\text{M}_\odot$ NS and a $10\,\text{M}_\odot$ BH and using the minimum $P_\text{b}$ to which each segment is sensitive, we derive limiting acceleration values using Kepler's third law, \begin{equation}
\left|a_\text{max}\right| = \left(\frac{2\pi}{P_\text{b}}\right)^{4/3}\left(\text{T}_\odot f\right)^{1/3}c,
\end{equation} 
where $c$ is the speed of light, $\text{T}_\odot$ is defined as
\begin{equation}\label{tdot}
\text{T}_\odot = \frac{G\text{M}_\odot}{c^{3}}=4.925490947\,\mu\text{s}
\end{equation}
where $G$ is Newton's gravitational constant, and $f$ is the mass function as defined by
\begin{equation}\label{eqn: mass function}
f = \frac{\left(m_\text{c}\sin i\right)^{3}}{\left(m_\text{c}+m_\text{p}\right)^{2}}.
\end{equation}
where $m_\text{p}$ and $m_\text{c}$ are the mass of the pulsar and the companion respectively, both in units of $\text{M}_\odot$. These acceleration ranges are also listed in Table~\ref{tab: segment parameters}. Meanwhile, the comparably-narrow acceleration search range adopted for the full-length $s=1$ segment is intended to optimise sensitivity to mildly-accelerated binary systems in wider orbits. Our choice of acceleration search step size for each segment is adopted from \cite{ekl+13}.

\subsection{Candidate selection and confirmation}\label{subsec: candidate selection}

In a `conventional' survey, pulsar candidates are typically selected for further inspection if their spectral signal-to-noise ratio ($\text{S/N}$) is greater than a threshold value of $\text{S/N}_\text{min}$. The appropriate value of $\text{S/N}_\text{min}$ can normally be derived using an assessment of the false-alarm statistics with respect to the number of dependent trials \citep[see e.g.][]{lk05}. However, in the case of the partially-coherent segmented acceleration search, this method is complicated by the multiple-pass, segmented nature of the search, which involves different numbers of both dependent and independent trials. As a conservative assessment, we have calculated the $\text{S/N}_\text{min}$ for each iteration of the segmented search ($s=1,2,4,8$), considering only the number of dependent trials. From this, we derive a lower limit of $\text{S/N}_\text{min}\simeq9.3$. We note that this value is likely to be an underestimate, as it fails to consider the complete search-space of the pipeline simultaneously, as well as the presence of RFI which will raise the survey's noise floor.

However, given that erring towards a lower value of $\text{S/N}_\text{min}$ is generally preferable, we conservatively consider each pulsar candidate produced by the pipeline with a spectral $\text{S/N}>8$. Each of these candidates is then folded, and those candidates whose folded $\text{S/N} > 8$ are then manually inspected by eye to assess the likelihood of the candidate representing a true pulsar discovery. Promising candidates are then reobserved with the Parkes 64-m Radio Telescope to confirm them as pulsars should they be redetected. Once confirmed, regular timing observations of each pulsar are conducted with a cadence of approximately one month, supplemented by intervals of higher cadence observations as required to obtain a phase-connected solution. Those pulsars with declinations  $\delta<-30^\circ$ are timed exclusively at Parkes, while those with $\delta>-30^\circ$ are typically passed to Jodrell Bank to be observed using the 76-m Lovell Radio Telescope.

Our confirmation strategy follows that of \cite{ncb15}, and involves reobserving each candidate using a `Ring-of-3' set of grids (labeled A-B-C) in a triangular configuration, each offset from the central discovery position (Grid D) by $\sim0.139^\circ$ such that they are separated from each other by the full width at half-maximum (FWHM) of the MB20 receiver ($0.24^\circ$). Typically, each of these grids is observed in turn until the pulsar is redetected, using a reduced $t_\text{int}$ designed to redetect the candidate pulsar at approximately $\text{S/N}=10$. If no redetection is made in any of the offset grids (A-B-C), additional confirmation observations may be taken at the discovery position (D). If available, archival data from the Parkes Multibeam Pulsar Survey \citep[PMPS;][]{mlc+01} as well as data from the HTRU-S Medium Latitude survey \citep[HTRU-S MedLat;][]{kjvs10} may also be searched to obtain additional information regarding the position and timing properties of the new pulsar.

\subsection{Status of survey processing}

Of the 1230 scheduled pointings which comprise HTRU-S LowLat, 536 ($\sim44\,\%$) have been processed through the partially-coherent segmented acceleration search as part of this paper. This is in addition to the 618 pointings ($\sim50\,\%$) which were previously processed by \cite{ncb15}, of which 180 ($\sim15\,\%$) have only been processed through a `standard', non-acceleration search \citep[for details, see][]{kjvs10}. A further 51 pointings encountered an error during processing, either due to data corruption or an error in the operation of the pipeline, in which case they may yet be recoverable by future processing efforts. In total, these 1205 pointings account for $\sim98\,\%$ of the entire HTRU-S LowLat pulsar survey\footnote{The pointings accounting for the remaining $2\,\%$ of the survey could not be processed as their data was unavailable for recall.}, with $\sim94\,\%$ having been successfully processed and reviewed.

\section{Known pulsar redetections}\label{sec: redetections}

In order to verify that the processing pipeline achieved the expected sensitivity, a complete record of the expected redetections of previously known pulsars in the survey region has been maintained, following the same procedure as outlined in \cite{ncb15}. For each survey beam, the current \textsc{psrcat} parameters of each nearby pulsar, including the pulsar's spin period $P$, its effective pulse width $W_\text{eff}$ (which we approximate as $W_{50}$, the width of the pulse at $50\,\%$ of its peak value) and its flux density at 1.4\,GHz, $S_{1400}$ were recorded. For those pulsars for which a $W_\text{50}$ is not recorded, a conservative value of $W_\text{50}$ was assumed. Due to the non-uniform response of each telescope beam, any offset of the pulsar from the center of the beam will cause a reduction in its apparent flux density, and correspondingly a reduction in its measured S/N. In order to account for this, we approximate the response pattern of the telescope beam as a Gaussian curve and calculate the expected apparent flux density,
\begin{equation}\label{eqn: flux position offset}
S_\text{exp} = S_{1400}e^{-\theta^2/2\sigma^2},
\end{equation} 
where $\theta$ is the offset in degrees and $\sigma$ is related to the FWHM of the telescope beam by
\begin{equation}\label{eqn: HTRU FWHM}
\sigma = \frac{\text{FWHM}}{2\sqrt{2\ln2}}.
\end{equation}
With each beam of the Parkes MB20 receiver having a FWHM of approximately $0.24^{\circ}$, this results in $\sigma\simeq0.1^\circ$. Based upon the modified value of $S_\text{exp}$, the expected S/N ($\text{S/N}_\text{exp}$) is then derived using the radiometer equation,
\begin{equation}\label{eqn: radiometer equation}
\text{S/N}_\text{exp} = \left(\frac{S_\text{exp}G\sqrt{n_\text{p}t_\text{int}\Delta f}}{\beta T_\text{sys}}\right)\left(\frac{P-W_{50}}{W_{50}}\right)^{1/2}
\end{equation}
where $G$ is the telescope gain in $\text{K}\,\text{Jy}^{-1}$ and varies between 0.735, 0.69 and 0.581 for the central, inner and outer beams of the receiver (see Table~3 of \citealt{kjvs10}). Meanwhile, the `degradation factor' $\beta\simeq1.16$, the effective bandwidth of the receiver $\Delta f=340\,\text{MHz}$, and the system temperature $T_\text{sys} = T_\text{sky}+T_\text{rcvr}$ where $T_\text{sky}\simeq7.6\,\text{K}$ (a mean value for the survey region, see Table~4 of \citealt{kjvs10}) and $T_\text{rcvr}\simeq{23}\,\text{K}$. For our observations, the number of polarisations $n_\text{p}=2$ and the integration time $t_\text{int}=4300\,\text{s}$.

We report \PSRindividualREDETECTIONS individual redetections of \PSRuniqueREDETECTIONS unique pulsars from the reported $44\,\%$ of the HTRU-S LowLat survey. A full record of these \PSRindividualREDETECTIONS redetections can be found in Appendix~\ref{ap-subsec: redetections}. Considering all HTRU-S LowLat survey data processed to date, this results in 1667 redetections spanning a combined total of 649 unique pulsars\footnote{This number is less than the sum of the totals from each portion of the survey due to the mutual overlap in the sets of individual pulsars detected in each half.}.

\begin{figure*}
 \begin{center}
\includegraphics[height=\textwidth, angle=270]{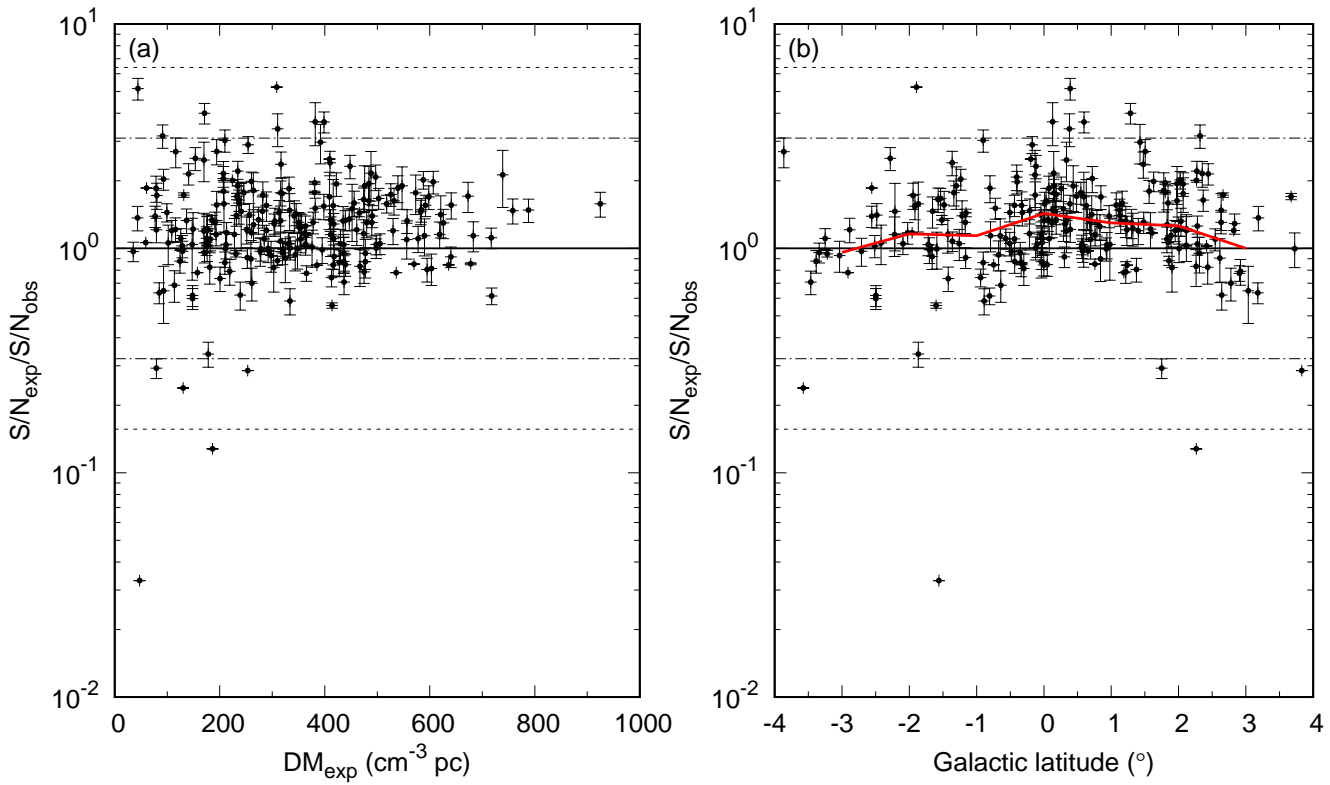}
 \end{center}
\caption{A comparison of the expected versus observed S/N values ($\text{S/N}_\text{exp}$ and $\text{S/N}_\text{obs}$ respectively) for pulsars redetected in $44\,\%$ of the HTRU-S LowLat survey. Panel (a) shows $\text{S/N}_\text{exp}/\text{S/N}_\text{obs}$ plotted against the \textsc{psrcat} DM for each redetected pulsar, while panel (b) shows $\text{S/N}_\text{exp}/\text{S/N}_\text{obs}$ plotted as a function of each pulsar's Galactic latitude. The bold line in each case represents the 1:1 relation, while the surrounding dashed lines extending outwards from the 1:1 line indicate contours containing $95\,\%$ and $99\,\%$ of the redetections respectively. The red line in panel (b) shows the median of the scatter, determined using bins with a width of $1^\circ$.}\label{fig: snr expected vs obs}
\end{figure*}

Figure~\ref{fig: snr expected vs obs} shows a comparison of the calculated $\text{S/N}_\text{exp}$ against the measured S/N ($\text{S/N}_\text{obs}$) for a subset of the 755 pulsar redetections reported here. As the response pattern of the telescope beam deviates from our Gaussian approximation outside of the beam FWHM, all redetections with an offset $\theta>\text{FWHM}/2=0.12^{\circ}$ (totaling 434 redetections) have been excluded from this comparison. Also excluded are two redetections of pulsars without a recorded $S_{1400}$, for which $\text{S/N}_\text{exp}$ cannot be calculated. Finally, 93 redetections are excluded due to the position of the relevant survey beams being sufficiently ambiguous to prevent an accurate determination of $\theta$ and hence allow for an accurate determination of $\text{S/N}_\text{exp}$. This ambiguity appears to have been caused by an error in the recorded position of each beam at the time the survey was taken, affecting the header information of the recorded files. In total, after accounting for these caveats, 226 pulsar redetections remain for the purposes of this analysis.

As shown in Figure~\ref{fig: snr expected vs obs}, a significant majority of redetections fall close to the 1:1 relation as expected. However, the division of data points around this relation does not appear to be symmetric, with 163 redetections (approximately $72\,\%$) having $\text{S/N}_\text{exp} > \text{S/N}_\text{obs}$. As noted by both \cite{kjvs10} and \cite{ncb15}, this is likely to be partly due to a reporting bias, where the highest values of $\text{S/N}$ observed during each known pulsar's initial set of observations tend to be reported (the variation in S/N between observations being due to scintillation and other potential instrumental effects, as further discussed in \citealt{lbb+13}). This leads to higher catalogue values of $S_{1400}$ which in turn leads to higher values of $\text{S/N}_\text{exp}$. The potential effect of scintillation can be seen in panel (a) of Figure~\ref{fig: snr expected vs obs}, which plots the ratio of $\text{S/N}_\text{exp}/\text{S/N}_\text{obs}$ as a function of catalogue DM. The scatter of redetections is seen to increase towards lower DM values where the effects of scintillation are likely to be most prominent \citep[see e.g.][]{sutton71,backer75}.

Another contributing factor to the observed scatter seen in Figure~\ref{fig: snr expected vs obs} is our choice of a constant value of $T_\text{sky}$ in calculating $\text{S/N}_\text{exp}$. In reality, the value of $T_\text{sky}$ typically increases with decreasing absolute Galactic latitude ($\left|b\right|$), which would cause an additional overestimation of $\text{S/N}_\text{exp}$ at the lowest values of $\left|b\right|$ \citep{hks+81}. This effect is clearly seen in panel (b) of Figure~\ref{fig: snr expected vs obs}, where the red line shows the median of the $\text{S/N}_\text{exp}/\text{S/N}_\text{obs}$ scatter as a function of $b$, clearly peaking close to $b=0^\circ$.

\subsection{Non-detections of known pulsars}\label{subsec: non detections}

We also note a number of non-detections of known pulsars expected to be detectable above a threshold folded S/N ($\text{S/N}_\text{min}$). For this analysis we set $\text{S/N}_\text{min}=9$ to maintain consistency with the previous work conducted in \cite{ncb15}. In addition, we again restrict our analysis to those non-detections with an offset $\theta\leq0.12^\circ$ and whose beam position is unambiguous.

Under these criteria, we identify 21 non-detections spanning 21 unique pulsars. We note that none of these 21 pulsars are known to be in binary systems. At present, we are able to account for 12 of the non-detections, with 9 remaining unexplained. Details of these non-detections are given in Appendix~\ref{ap-subsec: non detections}. 

Considering these non-detected pulsars in comparison to the set of redetected pulsars, we calculate that $\sim2\,\%$ of expected pulsars have been missed during the processing of the $\sim44\,\%$ of the HTRU-S LowLat survey data processed in this paper. This is comparable with the $\sim1\,\%$ non-detection rate reported for the $\sim50\,\%$ processed by \cite{ncb15}, indicating that the more-recently processed data has been analysed to an approximately equivalent sensitivity.

\subsection{Binary redetections}\label{subsec: binary redetections}

As the goal of the partially-coherent segmented acceleration search pipeline is to enhance our sensitivity to binary pulsars, we have also maintained a record of the observed S/N of each redetected binary pulsar across each searched segment. A total of 17 unique binary pulsars were detected across 28 individual survey beams. The highest S/N redetections and detected accelerations of each pulsar across all segments are provided in Appendix~\ref{ap-subsec: binary redetections}. As indicated by the listed values of $P_\text{b}$, the majority of these pulsars (up to and including PSR~J1431$-$5740) are of sufficiently-long orbital periods so as to be easily detectable without the need for a segmented acceleration search, and were typically detected at maximum S/N in the full-length observation.

More interesting behaviour is observed as $P_\text{b}$ shrinks to the point where $r_\text{orb}$ (as defined by Equation~\ref{eqn: rorb}) approaches $0.1$ for the full-length observation (corresponding to a critical orbital period defined as $P_\text{b,crit} =12\,\text{h}$). Here we consider the examples of both PSR~J1435$-$6100 ($P_\text{b} = 32.5\,\text{h}$) and PSR~J1802$-$2124 ($P_\text{b} = 16.8\,\text{h}$). Although the orbital period of both pulsars is larger than $P_\text{b,crit}$, the maximum line-of-sight accelerations of both pulsars exceeds the $\pm1\,\text{m\,s}^{-2}$ search range of the full-length segments (as listed in Table~\ref{tab: segment parameters}). This is likely a significant factor in the non-detection of either pulsar in the full-length segment, with each pulsar only being detected in shorter segments which were searched at larger acceleration values. This therefore appears to represent a parameter space to which our pipeline may not be sensitive. Pulsars with similar orbital parameters ($P_\text{b} > P_\text{b,crit}$ and $\left|a\right| > 1\,\text{m\,s}^{-2}$) but lower flux densities may not be detectable in the shorter segments, requiring the sensitivity of the full-length observation to be identified. However, as our search is specifically intended to target those pulsars for which $P_\text{b} < P_\text{b,crit}$, we do not consider this a great concern.

Additionally, the two beams in which PSR~J1802$-$2124 was detected (as listed in Table~\ref{ap-tab: binary redetections}) highlight the dependence of the search sensitivity on the orbital phase ($\varphi$) at which a given pulsar was observed. In beam 2011-12-30-23:14:07/02, PSR~J1802$-$2124 experienced accelerations low enough for a near-optimal detection in the full-length observation. However, both the full-length and half-length segment detections in this beam show the presence of jerk ($\dot{a}$), which is at its maximum magnitude in a circular orbit when $a\simeq0\,\text{m\,s}^{-2}$. It is at these orbital phases that the $r_\text{orb}\simeq0.1$ approximation is least applicable, with a smaller $r_\text{orb}$ being favoured \citep{ncb15}. This contributes to the higher S/N detection seen in the half-length segment of 2011-12-30-23:14:07/02. In contrast, the detection of PSR~J1802$-$2124 in beam 2011-10-12-04:24:15/07 occurs at an acceleration close to $a_\text{max}$ where the jerk $\dot{a}\simeq0$, and no evidence of jerk is seen in the half-segment detection.

Finally, PSR~J1141$-$6545 represents the only known short-$P_\text{b}$, relativistic binary that was observed and redetected during this portion of the survey processing\footnote{One additional short-$P_\text{b}$, relativistic binary (PSR~J1756-2251) exists within the HTRU-S LowLat survey region, but did not fall within any of the survey beams processed for this paper.}. The pulsar experiences high line-of-sight accelerations\footnote{The maximum and minimum line-of-sight accelerations of PSR~J1141$-$6545 change over time due its high rate of periastron advance of $\dot{\omega}\simeq5.31^\circ\,\text{yr}^{-1}$ \citep{bbv+08}.} and is also moderately eccentric with $e=0.17$, making it a unique test case for the segmented acceleration search pipeline. In all four beams in which the pulsar was detected, the maximum S/N detection occurred in the half-length segment, for which $r_\text{orb}=0.13$. With an $r_\text{orb}=0.06$, the quarter-length segments are also near-ideally suited to a detection of the pulsar, and had these segments been searched at smaller values of acceleration it is likely the detected S/N in these segments would have been consistent with the expected $\sqrt{2}$ reduction in S/N from the half-segment values. However, the stronger S/N of the half-segment detections indicates a successful application of the segmented-search strategy to a relativistic binary pulsar.

\section{Newly-discovered pulsars}\label{sec: new discoveries}

\begin{table*}
\begin{center}
\caption{For each of the \PSRnumADC newly-discovered pulsars from the HTRU-S LowLat survey, we list here the survey beam in which each pulsar was discovered (identified by its starting UTC time stamp and beam number), along with the S/N at which each pulsar was detected in the survey ($\text{S/N}_\text{HTRU}$) and, if available, the S/N of the pulsar in the PMPS ($\text{S/N}_\text{PMPS}$). Also listed are the mean flux density ($S_{1400}$), pulse widths at 50\,\% and 10\,\% of the pulse peak ($W_{50}$ and $W_{10}$) and derived luminosity ($L_{1400}$) of each pulsar. Values in parentheses, where available, represent 1-$\sigma$ uncertainties on the final digit. $L_{1400}$ is based on the DM distance estimates to each pulsar, using the NE2001 model \citep[left column;][]{NE2001a} and the YMW16 model \citep[right column;][]{YMW16}.}\label{tab: discovery summary}
\begin{tabular}{lcrrlccrr}
\hline
PSR name & Pointing/Beam & $\text{S/N}_\text{HTRU}$ & $\text{S/N}_\text{PMPS}$ & $S_{1400}$ & $W_{50}$ & $W_{10}$ & \multicolumn{2}{c}{$L_{1400}$} \\
 & & & & (mJy) & (ms) & (ms) & \multicolumn{2}{c}{(mJy\,kpc$^{2}$)} \\
\hline
J1136$-$6527 & 2012-02-18-20:27:49/12 & 11.8 & - & 0.14(2) & 19.6 & 35.8 & 1.6 & 0.5 \\
J1210$-$6322 & 2011-10-10-20:41:56/08 & 11.6 & $<7.0$ & 0.151(15) & 66.0 & 116.0 & 18.0 & 13.8 \\
J1223$-$5856 & 2012-01-19-13:30:00/04 & 34.9 & - & 0.377(12) & 63.3 & 85.5 & 10.6 & 8.6 \\
J1300$-$6602 & 2012-02-18-21:41:30/01 & 13.5 & $<7.0$ & 0.119(15) & 22.1 & 96.3 & 21.5 & 28.9 \\
J1344$-$5855 & 2011-12-28-17:24:43/04 & 15.4 & 8.2 & 0.138(10) & 13.0 & 23.8 & 5.8 & 7.9 \\
J1430$-$5712 & 2012-01-19-16:45:38/01 & 13.2 & - & 0.092(16) & 10.0 & 39.8 & 0.81 & 1.4 \\
J1434$-$5943 & 2011-12-27-16:36:22/08 & 13.3 & $<6.3$ & 0.17(2) & 42.9 & 55.6 & 0.94 & 1.7 \\
J1504$-$5659 & 2011-12-13-18:40:47/05 & 14.2 & 7.5 & 0.11(2) & 49.0 & 60.3 & 6.0 & 14.1 \\
J1507$-$5800 & 2012-04-10-11:32:06/09 & 11.1 & 8.2 & 0.20(3) & 8.70 & 64.5 & 7.5 & 7.0 \\ 
J1513$-$6013 & 2012-07-21-06:33:26/08 & 18.0 & 8.9 & 0.20(4) & 35.3 & 64.2 & 6.5 & 12.4 \\
J1514$-$5316 & 2011-12-21-23:02:36/02 & 10.2 & 9.4 & 0.147(18) & 5.16 & 16.1 & 0.1 & 0.1 \\
J1537$-$5312 & 2011-12-23-18:06:55/08 & 14.8 & 9.3 & 0.458(15) & 1.84 & 2.35 & 3.8 & 4.3\\ 
J1547$-$5709 & 2011-12-12-20:12:12/03 & 17.5 & - & 0.34(2) & 0.149 & 0.905 & 1.2 & 2.5 \\
J1557$-$5151 & 2011-12-08-04:31:54/12 & 17.2 & $<7.5$ & 0.310(18) & 28.9 & 63.5 & 23.3 & 13.0 \\
J1603$-$5312 & 2012-08-03-05:51:26/09 & 14.0 & 9.0  & 0.25(5) & 24.4 & 45.9 & 4.4 & 2.4 \\
J1612$-$5022 & 2012-07-24-09:40:05/05 & 12.0 & - & 0.23(3) & 15.2 & 50.4 & 7.2 & 3.7 \\
J1615$-$4958 & 2012-07-24-09:40:05/11 & 14.2 & $<7.7$ & 0.158(17) & 9.13 & 61.8 & 4.3 & 2.4 \\
J1618$-$4624 & 2012-04-01-13:59:20/07 & 15.1 & 7.8 & 0.273(13) & 0.291 & 1.02 & 1.5 & 2.5 \\ 
J1634$-$4229 & 2012-03-31-19:56:13/07 & 19.9 & - & 0.16(2) & 15.0 & 82.2 & 7.3 & 57.5\\
J1653$-$4105 & 2012-09-24-04:17:22/12 & 15.8 & 8.8 & 0.269(16) & 22.8 & 45.0 & 9.3 & 53.0 \\
J1653$-$45 & 2012-07-20-11:04:49/02 & 11.2 & - & - & 15.9 & 29.1 & - & - \\
J1704$-$3756 & 2011-12-23-20:33:36/05 & 11.3 & - & 0.134(15) & 11.7 & 21.3 & 4.7 & 31.1 \\
J1706$-$4434 & 2011-12-12-05:16:53/05 & 16.9 & - & 0.19(2) & 11.8 & 21.6 & 14.5 & 70.9 \\
J1719$-$3458 & 2012-04-13-15:31:21/01 & 13.7 & 10.2 & 0.20(2) & 21.0 & 25.9 & 10.2 & 53.3 \\
J1727$-$2951 & 2011-10-12-03:10:49/04 & 16.7 & - & 0.514(8) & 11.4 & 18.9 & 7.6 & 15.1 \\
J1731$-$33 & 2012-10-04-10:29:13/13 & 11.6 & - & - & 48.4 & 88.7 & - & - \\
J1734$-$2859 & 2011-12-07-03:42:23/10 & 10.4 & 6.6 & 0.13(2) & 8.72 & 26.5 & 3.2 & 17.1 \\
J1745$-$23 & 2012-12-09-23:10:30/01 & 14.3 & - & - & 0.660 & 1.22 & - & - \\
J1749$-$2146 & 2012-04-01-17:39:19/07 & 16.9 & 10.8 & - & 121.3 & 153.2 & - & - \\
J1753$-$28 & 2013-02-01-01:43:53/09 & 19.8 & 8.2 & - & 3.90 & 7.85 & - & - \\
J1757$-$1854 & 2012-04-12-16:27:35/03 & 13.3 & - & 0.25(4) & 0.705 & 1.80 & 13.7 & 96.0 \\
J1810$-$1709 & 2011-12-31-22:59:57/05 & 13.9 & - & 0.45(4) & 134.2 & 309.0 & 38.3 & 94.7 \\
J1812$-$15 & 2011-10-11-06:44:35/06 & 42.5 & - & - & 18.7 & 34.2 & - & - \\
J1812$-$20 & 2012-07-21-15:18:42/08 & 23.2 & 10.7 & - & 102.8 & 392.0 & - & - \\
J1822$-$0719 & 2012-08-03-11:58:41/07 & 11.4 & - & - & 10.1 & 18.5 & - & - \\
J1822$-$0902 & 2012-04-02-18:07:28/13 & 16.5 & 12.4 & - & 5.25 & 7.11 & - & - \\
J1831$-$04 & 2011-12-28-01:09:49/03 & 18.4 & - & - & 15.3 & 28.1 & - & - \\
J1835$-$0600 & 2011-10-11-07:58:09/13 & 16.2 & 8.6 & - & 29.0 & 53.6 & - & - \\
J1851$-$06$^{\text{a}}$ & 2012-08-04-14:47:04/06 & 9.5 & $<6.6$ & - & 50.2 & 63.9 & - & - \\
J1854$-$0524 & 2012-04-11-21:49:28/01 & 23.0 & - & - & 12.6 & 23.3 & - & - \\
\hline
\multicolumn{9}{l}{\footnotesize{$^{\text{a}}$ PSR~J1851$-$06 was later discovered independently during the commissioning of FAST and its pilot}} \\
\multicolumn{9}{l}{\footnotesize{observations for the Commensal Radio Astronomy FAST Survey \citep[CRAFTS;][]{lwq+18}.}} \\
\end{tabular}
\end{center}
\end{table*}

\begin{figure*}
 \begin{center}
 \scalebox{0.86}{
  \input{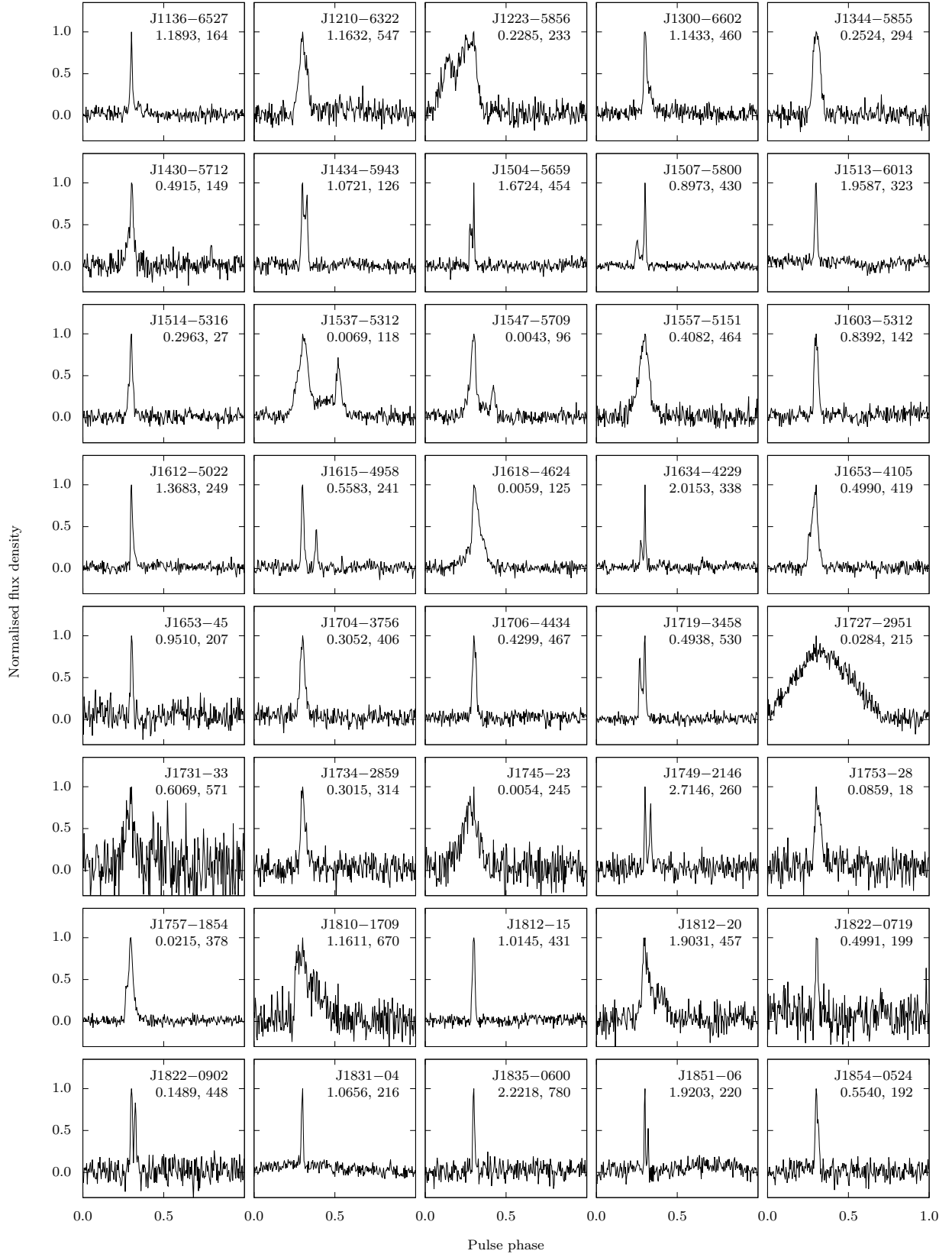}
 }
 \end{center}
\caption{Integrated pulse profiles of the \PSRnumADC newly-discovered pulsars. Each profile consists of 256 phase bins, has had its peak amplitude normalised to unity and has been rotated such that its peak is at a pulse phase of 0.3. Listed in the top right of each profile is the pulsar's current name, spin period (s) and DM ($\text{cm}^{-3}\,\text{pc}$).}\label{fig: discovery profiles}
\end{figure*}

A total of \PSRnumADC new pulsars\footnote{See footnote~\ref{ftnt: J1757-footnote}.} have been discovered in data processed for this paper (see Table~\ref{tab: discovery summary}). All but one of these pulsars have been successfully confirmed through reobservation with the Parkes 64-m Radio Telescope using the gridding strategy outlined in Section~\ref{subsec: candidate selection}. The remaining pulsar (PSR~J1831$-$04) displays evidence of nulling and/or intermittency in its discovery observation (see Section~\ref{subsec: long P0 binaries}) and is considered sufficiently unambiguous in this observation so as to be `self-confirmed'\footnote{A candidate can be considered \textit{`self-confirmed'} if it is detected with a high S/N (e.g. $> 15$), exhibits broadband and continuous emission, and has a DM constrained away from $0\,\text{cm}^{-3}\,\text{pc}$, such that the likelihood of it not representing a genuine pulsar detection is remote.}.

As of the time of writing, \PSRnumADCunsolved of the \PSRnumADC newly discovered pulsars lack sufficient pulse times-of-arrival (TOAs) for the determination of unique phase-connected timing solutions. These pulsars have been allocated temporary names listing only two digits of declination and are listed in Table~\ref{tab: unsolved pulsars}. The remaining \PSRnumADCsolved pulsars for which full timing solutions have been developed are listed in Tables~\ref{tab: solved pulsars main}, \ref{tab: solved pulsars supplementary} and \ref{tab: MSP-WD binary params} (with the exception of PSR~J1757$-$1854). Pulsars in these tables for which the uncertainty in declination is greater than or equal to $0.5'$ have also been assigned a temporary name with only two digits of declination.
\begin{table*}
\caption{Specifications of the telescope receiver and backend combinations used for timing observations, including the antenna gain ($G$), system temperature ($T_\text{sys}$), central observing frequency ($f_\text{c}$), and observing bandwidth ($B$).}\label{tab: Timing Setup}
\begin{center}
\begin{tabular}{lllllll}
  \hline
  Telescope & Receiver & $G$ & $T_\text{sys}$ & Backend & $f_\text{c}$ & $B$  \\
  & & ($\text{K\,Jy}^{-1}$) & (K) & & (MHz) & (MHz) \\
  \hline
  Parkes & MB20 & 0.74 & 23 & BPSR$^{\text{a}}$ & 1382 & 400 \\
  & & & & DFB4 & 1369 & 256 \\
  & & & & CASPSR$^{\text{a}}$ & 1382 & 400 \\
  & H-OH & 0.83 & 25 & DFB4 & 1369 & 256 \\
  & & & & CASPSR$^{\text{a}}$ & 1382 & 400 \\
  \hline
  Jodrell & L-band & 1.00 & 28 & DFB & 1532 & 384 \\
  & & & & ROACH & 1527 & 400 \\
  \hline
  \multicolumn{7}{l}{\footnotesize{$^{\text{a}}$ The usable bandwidth of BPSR and CASPSR is reduced to 340\,MHz}} \\
  \multicolumn{7}{l}{\footnotesize{due to the presence of strong RFI from the \textit{Thuraya 3} satellite and an}} \\
  \multicolumn{7}{l}{\footnotesize{associated RF filter.}} \\
\end{tabular}
\end{center}
\end{table*}

\begin{table*}
\begin{center}
\caption{Discovery parameters of the \PSRnumADCunsolved newly-discovered pulsars for which full timing solutions are not yet available. The reported spin period ($P$) and dispersion measure (DM) of each pulsar is taken from its discovery observation. The reported position of each pulsar in both equatorial (R.A. and Dec.) and galactic ($l$ and $b$) coordinates represents the best-known gridded position of the pulsar following confirmation observations. DM distances are calculated according to the NE2001 model \citep[left column;][]{NE2001a} and the YMW16 model \citep[right column;][]{YMW16}. Values in parentheses represent 1-$\sigma$ uncertainties on the final digit.}\label{tab: unsolved pulsars}
\begin{tabular}{lccrrllrr}
\hline
PSR name & R.A. (J2000) & Dec. (J2000) & $l$ & $b$ & $P$ & DM & \multicolumn{2}{c}{Dist.}\\
 & ($^\text{hms}$) & ($^{\circ\,'\,''}$) & ($^\circ$) & ($^\circ$) & (ms) & ($\text{cm}^{-3}\,\text{pc}$) & \multicolumn{2}{c}{kpc} \\
\hline
J1653$-$45$^{\text{a}}$ & 16:53.9(3) & $-$45:17(7) & 340.75 & $-$0.97 & 950.977(3) & 207(9) & 3.6 & 3.5 \\
J1731$-$33 & 17:31.8(4) & $-$33:48(7) & 354.31 & $-$0.11 & 606.9003(16) & 571(9) & 6.1 & 4.3 \\
J1745$-$23$^{\text{a}}$ & 17:45.5(4) & $-$23:25(7) & 4.70 & 2.89 & 5.41669986(14) & 244.94(9) & 4.5 & 7.9 \\
J1753$-$28 & 17:53.1(4) & $-$28:53(7) & 0.89 & $-$1.38 & 85.85861(2) & 18.0(9) & 0.6 & 0.7 \\
J1812$-$15$^{\text{a}}$ & 18:12.6(5) & $-$15:31(7) & 14.74 & 1.29 & 1014.529(3) & 431(10) & 5.9 & 10.0 \\ 
J1831$-$04$^{\text{b}}$ & 18:31.0(5) & $-$04:29(7) & 26.62 & 2.49 & 1065.578(3) & 216(10) & 4.4 & 4.9 \\
J1851$-$06 & 18:51.2(5) & $-$06:38(7) & 27.08 & $-$3.02 & 1920.312(13) & 220(20) & 4.8 & 5.7 \\
\hline
\multicolumn{9}{l}{\footnotesize{$^{\text{a}}$ Indicates confirmed binary pulsars.}} \\
\multicolumn{9}{l}{\footnotesize{$^{\text{b}}$ Indicates candidate binary pulsars.}} \\
\end{tabular}
\end{center}
\end{table*}

Timing observations for each pulsar were conducted by the Parkes 64-m Radio Telescope and the Jodrell Bank 76-m Lovell Radio Telescope. At Parkes observations were conducted with an approximately monthly cadence. Jodrell Bank timing observations were conducted with an irregular cadence, with observations made of each pulsar typically every one to three weeks. Parkes observations were conducted using two receivers, the MB20 receiver and the H-OH 21-cm receiver\footnote{The H-OH receiver was used between MJD~57440 and 57727 due to the unavailability of the MB20 receiver.}. The two timing backends employed at Parkes include a Digital Filter Bank backend system (DFB4), capable only of incoherent dedispersion, and the CASPER Parkes Swinburne Recorder\footnote{http://www.astronomy.swin.edu.au/pulsar/?topic=caspsr} (CASPSR), capable of coherent dedispersion. Additionally, search-mode filterbank data taken using the Berkeley Parkes Swinburne Recorder\footnote{http://www.astronomy.swin.edu.au/pulsar/?topic=bpsr} (BPSR) to HTRU specifications \citep[see][]{kjvs10} was also used in the early timing stages of multiple pulsars. Jodrell Bank observations were conducted using the single-pixel L-band receiver in combination with both a DFB backend and a ROACH\footnote{Based on the ROACH FPGA processing board, see https://casper.berkeley.edu/wiki/ROACH} backend \citep{bjk+16} capable of coherent dedispersion. A summary of the receivers and backends used in this project\footnote{For the timing specifications of PSR~J1757$-$1854, refer to \cite{cck+18}.} is presented in Table~\ref{tab: Timing Setup}.

Each timing solution was determined using multiple software packages including the \textsc{dspsr}\footnote{https://sourceforge.net/projects/dspsr} \citep{dspsr}, \textsc{psrchive}\footnote{http://psrchive.sourceforge.net} \citep{hvsm04}, \textsc{sigproc}\footnote{http://sigproc.sourceforge.net} \citep{sigproc} and \textsc{presto}\footnote{http://www.cv.nrao.edu/~sransom/presto} \citep{ransom01} pulsar data analysis packages as well as the \textsc{tempo}\footnote{http://tempo.sourceforge.net} and \textsc{tempo2}\footnote{http://www.atnf.csiro.au/research/pulsar/tempo2} \citep{hem06} timing software packages. Each observation was first cleaned to remove instances of time and frequency-domain RFI, and then calibrated against an observation of a pulsed noise diode to account for the differential gain and phase between the receiver's polarisation feeds. TOAs were then produced by summing each observation in both frequency and polarisation, before partially summing in time and cross-correlating each summed profile against a standard reference pulse profile. Initial timing solutions were typically developed using the \textsc{tempo} software package, often with the use of a modified prototype version of the \textsc{dracula}\footnote{https://github.com/pfreire163/Dracula} software package, which solves for the global rotation count of a pulsar between discrete observations using the phase-jump technique described by \cite{dracula}. The finalised timing solutions presented in Tables~\ref{tab: solved pulsars main} and \ref{tab: solved pulsars supplementary} were produced using \textsc{tempo2}, after first reweighting each set of TOAs such that their reduced $\chi^2=1$. All solutions in Tables~\ref{tab: solved pulsars main} and \ref{tab: solved pulsars supplementary} are in TCB\footnote{Barycentric Coordinate Time} units and use the DE421\footnote{https://ssd.jpl.nasa.gov/?ephemerides\#planets} planetary ephemeris.

\begin{table*}
\begin{center}
\caption{Best-fit \textsc{tempo2} timing parameters for \PSRnumADCsolvedminusbinaries pulsars, including their positions in both equatorial (R.A. and Dec.) and galactic ($l$ and $b$) coordinates, spin periods ($P$), spin-period derivatives ($\dot{P}$) and dispersion measures (DM). Values in parentheses represent 1-$\sigma$ uncertainties on the final digit. DM distances are calculated according to the NE2001 model \citep[left column;][]{NE2001a} and the YMW16 model \citep[right column;][]{YMW16}.}\label{tab: solved pulsars main}
\begin{tabular}{lllrrlllrr}
\hline
PSR name & R.A. (J2000) & Dec. (J2000) & $l$ & $b$ & $P$ & $\dot{P}$ & DM & \multicolumn{2}{c}{Dist.} \\
 & ($^\text{hms}$) & ($^{\circ\,'\,''}$) & ($^\circ$) & ($^\circ$) & (ms) & ($10^{-18}$) & ($\text{cm}^{-3}\,\text{pc}$) & \multicolumn{2}{c}{(kpc)} \\
\hline
J1136$-$6527 & 11:36:25.22(7) & $-$65:27:19.5(3) & 295.24 & $-$3.71 & 1189.30900495(8) & 1750(16) & 164.1(18) & 3.4 & 2.0 \\
J1210$-$6322 & 12:10:46.78(10) & $-$63:22:20.9(6) & 298.40 & $-$0.86 & 1163.18571439(18) & 9247(12) & 547(5) & 10.9 & 9.6 \\
J1223$-$5856 & 12:23:40.14(3) & $-$58:56:48.9(2) & 229.35 & 3.73 & 288.54142685(2) & 5.7(8) & 233(3) & 5.3 & 4.8 \\
J1300$-$6602 & 13:00:26.93(3) & $-$66:02:16.8(2) & 303.85 & $-$3.18 & 1143.31584211(4) & 335(2) & 460(3) & 13.4 & 15.6 \\
J1344$-$5855 & 13:44:53.05(3) & $-$58:55:24.77(18) & 309.79 & 3.23 & 252.397929468(5) & 2914.5(17) & 294.1(6) & 6.5 & 7.5 \\
J1430$-$5712 & 14:30:16.17(2) & $-$57:12:31.4(2) & 316.04 & 3.13 & 491.518803580(9) & 46940.8(16) & 149(4) & 3.0 & 3.9 \\
J1434$-$5943 & 14:34:58.31(2) & $-$59:43:41.6(2) & 315.65 & 0.56 & 1072.12121207(3) & 29.6(18) & 126(3) & 2.4 & 3.2 \\
J1504$-$5659 & 15:04:31.987(17) & $-$56:59:19.4(2) & 320.33 & 1.32 & 1672.37234272(3) & 1428(3) & 454.2(11) & 7.3 & 11.3 \\
J1507$-$5800 & 15:07:03.504(8) & $-$58:00:56.20(10) & 320.12 & 0.26 & 897.254102404(11) & 260.3(6) & 429.5(5) & 6.1 & 5.9 \\
J1513$-$6013 & 15:13:54.02(4) & $-$60:13:26.5(3) & 319.75 & $-$2.09 & 1958.73704232(12) & 1371(8) & 322.7(16) & 5.7 & 7.9 \\
J1514$-$5316 & 15:14:40.160(11) & $-$53:16:02.3(2) & 323.46 & 3.80 & 296.279212139(4) & 1.6(5) & 27.1(3) & 0.9 & 0.9 \\
J1557$-$5151 & 15:57:29.30(2) & $-$51:51:08.4(3) & 329.56 & 1.13 & 408.154708451(8) & 75.4(9) & 464(3) & 8.7 & 6.5 \\
J1603$-$5312 & 16:03:50.88(3) & $-$53:12:58.0(4) & 329.40 & $-$0.53 & 839.22081265(6) & 49600(5) & 142(3) & 4.2 & 3.1 \\
J1612$-$5022 & 16:12:28.028(12) & $-$50:22:57.3(4) & 332.30 & 0.66 & 1368.28292337(6) & 34(4) & 248.7(15) & 5.5 & 4.0 \\
J1615$-$4958 & 16:15:17.38(3) & $-$49:58:02.0(4) & 332.91 & 0.65 & 558.25750561(3) & 1461(8) & 240.7(6) & 5.2 & 3.9 \\
J1634$-$4229 & 16:34:14.665(9) & $-$42:29:44.3(4) & 340.54 & 3.54 & 2015.26299651(4) & 8010(2) & 337.9(10) & 6.7 & 18.9 \\
J1653$-$4105 & 16:53:25.374(10) & $-$41:05:25.6(4) & 343.95 & 1.76 & 498.978065960(7) & 54.1(11) & 419.4(12) & 5.9 & 14.0 \\
J1704$-$3756 & 17:04:57.466(15) & $-$37:56:42.9(8) & 347.80 & 1.95 & 305.234449799(11) & 11284(3) & 405.7(6) & 5.9 & 15.2 \\
J1706$-$4434 & 17:06:23.183(9) & $-$44:34:30.0(2) & 342.67 & $-$2.26 & 429.922423250(6) & 2578(2) & 467.0(6) & 8.8 & 19.4 \\
J1719$-$3458 & 17:19:12.141(3) & $-$34:58:22.4(2) & 351.89 & 1.39 & 493.774733755(3) & 14.9(3) & 530.0(3) & 7.2 & 16.4 \\
J1734$-$2859 & 17:34:00.114(10) & $-$28:59:53.4(16) & 358.60 & 2.12 & 301.455877926(9) & 8.0(6) & 313.9(9) & 4.9 & 11.4 \\
J1749$-$2146 & 17:49:21.241(19) & $-$21:46:31(10) & 6.57 & 2.98 & 2714.55556146(12) & 7077(7) & 260(20) & 4.8 & 9.6 \\
J1810$-$1709 & 18:10:28.21(3) & $-$17:09:27(4) & 13.05 & 0.96 & 1161.13257983(7) & 333(6) & 670(20) & 9.3 & 14.6 \\
J1812$-$20 & 18:12:36.58(5) & $-$20:58.1(5) & 9.95 & $-$1.32 & 1903.1119816(5) & 240(30) & 457(19) & 6.7 & 11.3 \\
J1822$-$0719 & 18:22:28.073(15) & $-$07:19:55.0(7) & 23.10 & 3.04 & 499.07553362(4) & 19(2) & 199(5) & 4.2 & 4.6 \\
J1822$-$0902 & 18:22:35.745(8) & $-$09:02:59.1(5) & 21.59 & 2.21 & 148.894507003(3) & 17780.9(8) & 448.1(15) & 7.9 & 15.7 \\
J1835$-$0600 & 18:35:20.25(5) & $-$06:00:00(4) & 25.76 & 0.83 & 2221.7871479(14) & 8430(80) & 780(20) & 9.8 & 10.6 \\
J1854$-$0524 & 18:54:55.35(2) & $-$05:24:23.1(19) & 28.51 & $-$3.24 & 544.02080981(15) & 1200(10) & 192(3) & 4.5 & 5.0 \\
\hline
\end{tabular}
\end{center}
\end{table*}

\begin{table*}
\begin{center}
\caption{Fit-related timing parameters for \PSRnumADCsolvedminusbinaries pulsars, including the data span, the period reference epoch, the number of TOAs used to derive each timing solution as listed in Table~\ref{tab: solved pulsars main}, the root mean square (RMS) of the weighted \textsc{tempo2} fit and the unweighted reduced $\chi^2$ ($\chi^2_\text{red}$). Derived parameters, including the characteristic age ($\tau_\text{c}$), the surface magnetic field ($B_\text{surf}$) and the spin-down luminosity ($\dot{E}$) are calculated according to the equations presented in \citet{lk05} and are given to a precision consistent with the measurements of $P$ and $\dot{P}$ used to derive them, up to a maximum of three significant figures.}\label{tab: solved pulsars supplementary}
\begin{tabular}{llllrrlll}
\hline
PSR name & Data span & Epoch & $n_\text{TOA}$ & RMS & $\chi^{2}_\text{red}$ & $\tau_\text{c}$ & $B_\text{surf}$ & $\dot{E}$ \\
 & (MJD) & (MJD) & & ($\mu\text{s}$) & & (Myr) & ($10^{10}\,\text{G}$) & ($10^{30}\,\text{erg\,s}^{-1}$) \\
\hline
J1136$-$6527 & 57732$-$58152 & 57895 & 28 & 1559 & 1.2 & 10.7 & 144 & 41.1 \\
J1210$-$6322 & 57530$-$58057 & 57648 & 38 & 4693 & 1.4 & 1.99 & 328 & 232 \\
J1223$-$5856 & 57372$-$58058 & 57405 & 46 & 2074 & 1.4 & 800 & 4.1 & 9 \\
J1300$-$6602 & 57372$-$58128 & 57554 & 46 & 1659 & 0.7 & 53.9 & 61.9 & 8.85 \\
J1344$-$5855 & 57732$-$58152 & 57942 & 43 & 1152 & 1.4 & 1.37 & 85.8 & 7160 \\
J1430$-$5712 & 57324$-$57845 & 57562 & 28 & 984 & 1.0 & 0.165 & 480 & 15600 \\
J1434$-$5943 & 57296$-$58058 & 57530 & 54 & 1628 & 1.6 & 570 & 17.8 & 0.95 \\
J1504$-$5659 & 57229$-$57845 & 57483 & 35 & 1172 & 1.7 & 18.5 & 155 & 12.1 \\
J1507$-$5800 & 57442$-$58127 & 57604 & 40 & 502 & 0.9 & 54.5 & 48.3 & 14.2 \\
J1513$-$6013 & 57500$-$58057 & 57633 & 55 & 2549 & 1.4 & 22.6 & 164 & 7.21 \\
J1514$-$5316 & 57837$-$58337 & 57892 & 46 & 561 & 1.8 & 2900 & 2.2 & 2 \\
J1557$-$5151 & 57066$-$57845 & 57359 & 36 & 1488 & 0.6 & 85.5 & 17.5 & 43.8 \\
J1603$-$5312 & 57586$-$58057 & 57693 & 48 & 1819 & 1.0 & 0.267 & 645 & 3310 \\
J1612$-$5022 & 57837$-$58337 & 57947 & 50 & 1116 & 0.9 & 640 & 21.6 & 0.52 \\
J1615$-$4958 & 57798$-$58128 & 57928 & 33 & 653 & 0.9 & 6.04 & 90.3 & 332 \\
J1634$-$4229 & 57442$-$58128 & 57603 & 27 & 656 & 0.5 & 3.98 & 402 & 38.7 \\
J1653$-$4105 & 57613$-$58153 & 57883 & 66 & 1484 & 1.0 & 146 & 16.4 & 17.2 \\
J1704$-$3756 & 57500$-$57866 & 57650 & 33 & 822 & 1.0 & 0.427 & 186 & 15700 \\
J1706$-$4434 & 57231$-$58126 & 57416 & 57 & 653 & 1.3 & 2.63 & 105 & 1280 \\
J1719$-$3458 & 57473$-$58059 & 57637 & 55 & 421 & 1.1 & 520 & 8.58 & 4.9 \\
J1734$-$2859 & 57500$-$58117 & 57674 & 27 & 673 & 0.9 & 600 & 4.91 & 12 \\
J1749$-$2146 & 57513$-$58115 & 57656 & 19 & 1213 & 1.1 & 6.06 & 438 & 14.0 \\
J1810$-$1709 & 57357$-$58123 & 57573 & 26 & 2612 & 1.3 & 55 & 62.2 & 8.4 \\
J1812$-$20 & 57512$-$58077 & 57615 & 27 & 3976 & 0.6 & 130 & 68 & 1.4 \\
J1822$-$0719 & 57657$-$58137 & 57700 & 21 & 836 & 1.5 & 420 & 9.7 & 6.0 \\
J1822$-$0902 & 57420$-$58116 & 57566 & 49 & 952 & 84.7 & 0.132 & 163 & 213000 \\
J1835$-$0600 & 57716$-$58126 & 57733 & 18 & 1714 & 1.2 & 4.16 & 433 & 30.4 \\
J1854$-$0524 & 57714$-$58149 & 57737 & 48 & 2288 & 13.2 & 7.16 & 80.8 & 294 \\
\hline
\end{tabular}
\end{center}
\end{table*}

For the \PSRnumADCdfbpks pulsars observed using the Parkes DFB4 backend and with full phase-connected timing solutions, it is also possible to determine calibrated 1.4-GHz flux densities ($S_{1400}$). Following polarisation calibration, each observation was flux calibrated against an observation of Hydra~A\footnote{Recorded as part of the Parkes Pulsar Timing Array (PPTA) project \citep[see e.g.][]{hydra_A}.}, which was typically separated in time from the pulsar observation by as much as one to two weeks. The phase-connected timing solution of each pulsar was then used to sum together the DFB4 observations so as to produce an integrated observation from which a measurement of flux density was derived. However, as the observed position of the pulsar is typically offset from the final timed position of the pulsar, an additional flux-density correction based on the offset in position ($\theta$) was applied using Equation~\ref{eqn: flux position offset}. The resulting values of $S_{1400}$ are listed in Table~\ref{tab: discovery summary}. Also derived are the 1.4-GHz luminosities of each pulsar, $L_{1400} = S_{1400}\times d^2$, where $d$ is the distance of each pulsar in kpc. Using the DM of each pulsar, a DM distance $d$ can be estimated based upon two separate models of the Galactic distribution of electron density, the NE2001 model \citep{NE2001a} and the YMW16 model \citep{YMW16}. These distance estimates are listed both in Table~\ref{tab: unsolved pulsars} and Table~\ref{tab: solved pulsars main}, while the corresponding values of $L_{1400}$ are listed in Table~\ref{tab: discovery summary}.

Figure~\ref{fig: discovery profiles} shows the integrated 1.4-GHz pulse profiles for each of the newly discovered pulsars, folded with 256 phase bins. In the case of the \PSRnumADCdfbpks pulsars timed using Parkes and with full phase-connected timing solutions, each integrated profile was produced by coherently summing the pulsar's timing observations. The integrated profiles of PSRs~J1537$-$5312, J1547$-$5709, J1618$-$4624, J1727$-$2951 and J1757$-$1854 (for which the intrachannel dispersion smearing of DFB4 becomes a significant factor) were produced using coherently dedispersed CASPSR data, while the remaining \PSRnumADCdfbpksminusmsp pulse profiles were produced from DFB4 data. Profiles for the \PSRnumADCunsummed pulsars without Parkes data or without phase-connected timing solutions were produced using either their discovery HTRU observation or their subsequent confirmation observation, whichever resulted in a higher value of S/N. All pulse profiles have had their baselines subtracted and peak amplitudes normalised to unity, and have been rotated such that the profile peak is located at a pulse phase of 0.3. For each integrated profile, an analytic model profile consisting of multiple Gaussian components was derived using the \textsc{psrchive} application \textsc{paas} and used to measure the pulse widths at 10\,\% ($W_{10}$) and 50\,\% ($W_{50}$) of the profile peak. These pulse widths are listed in Table~\ref{tab: discovery summary}.

Of the \PSRnumADC newly-discovered pulsars reported in this paper, \PSRnumADCconfbin have been conclusively determined to be in binary systems. The phase-connected timing solutions of four of these pulsars (PSRs~J1537$-$5312, J1547$-$5709,  J1618$-$4624 and J1727$-$2951) are presented in Table~\ref{tab: MSP-WD binary params}, while the solution and properties of PSR~J1757$-$1854 have been previously published in \cite{cck+18}. The remaining three confirmed binary pulsars (PSRs~J1653$-$45, J1745$-$23, and J1812$-$15) are highlighted in Table~\ref{tab: unsolved pulsars}. An additional pulsar also displays evidence which strongly suggests that it also represents a new binary system; this pulsar (PSR~J1831$-$04) is also presented in Table~\ref{tab: unsolved pulsars}. All of these binary systems are discussed in detail in Section~\ref{sec: pulsars of interest}.

\subsection{Redetections in the PMPS}

The HTRU-S LowLat survey area has a complete overlap with the survey area of the PMPS. A comprehensive search of PMPS archival data was therefore carried out to determine if any of the newly-discovered pulsars reported in this paper were detectable in the earlier survey. All PMPS beams within one beamwidth ($\sim0.24^\circ$) of the best known position of each pulsar were searched for matching pulsar candidates. For those pulsars with full timing solutions, a direct ephemeris fold of each observation was also produced and inspected. For those pulsars for which a detection was made, the maximum S/N derived from all inspected PMPS beams ($\text{S/N}_\text{PMPS}$) is recorded in Table~\ref{tab: discovery summary}.

In total, \PSRnumADCPMPStotal pulsars were detected in the archival PMPS data. Of these, \PSRnumADCPMPSovereight were detected above the theoretical S/N cutoff for pulsars in the PMPS, $\text{S/N}_\text{min,PMPS} = 8$ \citep{mlc+01}. Therefore, these pulsars were theoretically detectable in the earlier survey, but for reasons unknown were overlooked, potentially as a result of the large number of pulsar candidates produced during the processing of the PMPS or due to the presence of RFI. It should be noted that many of the pulsars detected in the PMPS using an ephemeris fold were not detected by a simple application of standard searching techniques (e.g. an FFT-based search). This difficulty in detection might also partially account for the previous non-discovery of these pulsars, and may allow for further PMPS detections in the future as additional timing solutions are derived. The remaining \PSRnumADCPMPSundereight pulsars were detected only weakly in the PMPS, falling below the $\text{S/N}_\text{min,PMPS}$ cutoff. In cases where the PMPS detection is only tentative, the measured $\text{S/N}_\text{PMPS}$ is set as an upper limit.

\subsection{Gamma-ray and supernova remnant associations}

We note at the outset that the chances of determining an association between any of the newly-discovered pulsars and a supernova remnant (SNR) would appear to be unlikely. SNRs have typical lifetimes of only $10^{4}\,\text{yr}$ \citep{fgw94}, while the youngest pulsars among our discoveries have characteristic ages on the order of $10^{5}\,\text{yr}$. While a SNR association is not impossible for pulsars in this age range, the lack of a well-measured proper motion for any of the discovered pulsars makes confirming any such association (should it exist) all the more difficult.

Nevertheless, we have attempted to evaluate the possibility of a SNR association for the 4 youngest pulsars (PSRs~J1430$-$5712, J1603$-$5312, J1704$-$3756 and J1822$-$0902), each of which has a characteristic age $\tau_\text{c}<1\,\text{Myr}$. A radius of association ($r_\text{A}$) was determined for each pulsar by taking the distribution of pulsars (as listed by \textsc{psrcat}) with constrained proper motions and $\tau_\text{c}<1\,\text{Myr}$. Using DM-distance estimates from both the NE2001 and YMW16 models, these proper motions were then converted into tangential velocities ($V_\text{T}$), of which the median value was taken. Using the characteristic age and DM-distance estimates of each of the newly-discovered pulsars (as listed in Tables~\ref{tab: solved pulsars main} and \ref{tab: solved pulsars supplementary}), these median $V_\text{T}$ values were converted into angular offsets, of which the larger was taken as $r_\text{A}$. The pulsars were then cross-matched against the Green SNR Catalogue \citep{green2017} for SNRs falling within these radii. 

For PSRs~J1430$-$5712, J1704$-$3756 and J1822$-$0902 (with $r_\text{A}$ values of $0.89^\circ$, $1.17^\circ$ and $0.27^\circ$ respectively), no SNR were identified. However, as these systems may have higher proper motions than assumed in our calculations, an undetected SNR association cannot be ruled out. Meanwhile, PSR~J1603$-$5312 (with an $r_\text{A}$ of $1.29^\circ$) was found to coincide with two SNR, namely G328.4$+$00.2 and G329.7$+$00.4 (separated by $1.25^\circ$ and $0.99^\circ$ respectively). In the case of G328.4$+$00.2, an association with PSR~J1603$-$5312 seems unlikely, as \cite{ggs+07} argue an age for the SNR of only $\sim6500\,\text{yr}$, approximately 40 times smaller than the pulsar's characteristic age. Furthermore, they argue that G328.4$+$00.2 is in fact a pulsar wind nebula (PWN) with its own embedded, as-yet undetected NS. In short, these facts would rule out any association between G328.4$+$00.2 and PSR~J1603$-$5312. No age has been determined for G329.7$+$00.4, and at present an association with PSR~J1603$-$5312 can neither be confirmed nor ruled out.

We have also not identified any gamma-ray associations for any of the newly-discovered HTRU-S LowLat pulsars. This was determined by conducting a search of $\sim9.6\,\text{yr}$ of data recorded by the Large Area Telescope (LAT) on the \textit{Fermi Gamma-Ray Space Telescope}. For each of the \PSRnumADCsolved pulsars with a phase-connected timing solution, a search was conducted using its best ephemeris, with details presented in \cite{sbc+18}. The lack of gamma-ray detections is not surprising, given that only one of the newly-discovered pulsars (PSR~J1430$-$5712) has an $\sqrt{\dot{E}}/d^{2}>10^{16}\,\left(\text{erg}\,\text{s}^{-1}\right)^{1/2}\,\text{kpc}^{-2}$, a typical criterion for gamma-ray pulsars \citep{aaa+13}. With additional timing, it is possible that the remaining unsolved pulsars may be shown to be associated with gamma-ray sources. 

\section{Individual pulsars of interest}\label{sec: pulsars of interest}

In addition to the previously-published PSR~J1757$-$1854, several of the pulsars discovered as part of this work warrant additional scientific scrutiny. They are discussed in the subsections below.

\subsection{PSR~J1537--5312 and PSR~J1547--5709, a pair of He-WD binary MSPs}\label{subsec: HE-WD MSPs}

PSRs~J1537$-$5312 and J1547$-$5709 are a pair of binary MSPs whose parameters are listed in Table~\ref{tab: MSP-WD binary params}. Based on their respective spin periods of $6.93\,\text{ms}$ and $4.29\,\text{ms}$ as well as their low spin-period derivatives ($\sim10^{-20}$ to $10^{-21}$), both of these pulsars appear to be highly recycled. The highly-circularised orbits and the range of companion masses implied by their mass functions (with minimum masses of $m_\text{c}=0.115\,\text{M}_\odot$ and $m_\text{c}=0.175\,\text{M}_\odot$ respectively) suggest that each pulsar possesses a degenerate helium white dwarf (He-WD) companion. These systems most-likely formed out of low-mass X-ray binaries (LMXB) through Case B Roche-lobe overflow (RLO) \citep[see e.g.][]{tauris11}. No known counterpart to these companion WDs can be found in the SIMBAD astronomical database\footnote{http://simbad.u-strasbg.fr/simbad/} \citep[\textsc{simbad};][]{simbad} to within a radius of 30 arcseconds, and no further confirmation observations of the WDs have been attempted. Such identifications may prove difficult given the large estimated distance to both pulsars ($\sim2-3\,\text{kpc}$) along with their positions in the interstellar medium-dense Galactic plane.

\begin{table*}
\begin{center}
\caption{Best-fit \textsc{tempo2} timing parameters for 4 newly-discovered binary pulsars. Values in parentheses represent 1-$\sigma$ uncertainties on the final digit. Values quoted as upper limits represent an uncertainty of 3\,$\sigma$. DM distances are calculated according to the NE2001 model \citep{NE2001a} and the YMW16 model \citep{YMW16}.}\label{tab: MSP-WD binary params}
\begin{tabular}{lllll}
\hline
Parameter & J1537$-$5312 & J1547$-$5709 & J1618$-$4624 & J1727$-$2951 \\
\hline
Right ascension, $\alpha$ (J2000) & 15:37:37.69466(18) & 15:47:24.1248(12) & 16:18:52.77579(13) & 17:27:00.402(5) \\
Declination, $\delta$ (J2000) & $-$53:12:25.057(3) & $-$57:09:17.5699(16) & $-$46:24:34.950(4) & $-$29:51:40.8(8) \\
Spin period, $P$ (ms) & 6.9270955083835(15) & 4.2911460641714(6) & 5.9313674952810(16) & 28.4049534402(6) \\
Spin period derivative, $\dot{P}$ ($10^{-18}$) & 0.01586(11) & 0.00745(3) & 0.00310(10) & 0.31(6) \\
Dispersion measure, DM ($\text{cm}^{-3}\,\text{pc}$) & 117.52(5) & 95.727(8) & 125.364(16) & 215.1(3) \\
\\
Binary model & ELL1 & ELL1 & ELL1 & ELL1 \\
Orbital period, $P_\text{b}$ (d) & 3.55014838(2) & 3.077476982(5) & 1.780433535(2) & 0.3951890(4) \\
Projected semi-major axis, $x$ (lt-s) & 1.982433(4) & 2.668161(2) & 5.329375(4) & 0.05783(11) \\
Epoch of the ascending node, $T_\text{asc}$ (MJD) & 57295.061729(3) & 57297.5850363(6) & 57560.5896560(4) & 57845.1434(2) \\
$\left|e\sin{\omega}\right|$, $\left|\epsilon_{1}\right|$ ($10^{-4}$) & $<0.149$ & $<0.0576$ & $<0.0444$ & $<130$ \\
$\left|e\cos{\omega}\right|$, $\left|\epsilon_{2}\right|$ ($10^{-4}$) & $<0.107$ & $<0.0431$ & $<0.0355$ & $<147$ \\
Inferred eccentricity, $e$ ($10^{-6}$) & $<11.9$ & $<4.75$ & $<4.27$ & $<14200$ \\
Mass function, $f$ ($10^{-3}\,\text{M}_\odot$) & 0.663721(4) & 2.153422(5) & 51.26961(11) & 0.001330(7) \\
Minimum companion mass$^\text{a}$, $m_\text{c,min}$ ($\text{M}_\odot$) & 0.1150667(2) & 0.17479358(15) & 0.5871916(5) & 0.01385(2) \\
Median companion mass$^\text{b}$, $m_\text{c,med}$ ($\text{M}_\odot$) & 0.1339705(3) & 0.20435198(17) & 0.7044464(6) & 0.01601(3) \\
\\
First TOA (MJD) & 57259 & 57091 & 57404 & 57839 \\
Last TOA (MJD) & 58136 & 58136 & 58136 & 58299\\
Timing epoch (MJD) & 57575 & 57445 & 57637 & 57958 \\
Number of TOAs, $n_\text{TOA}$ & 92 & 90 & 98 & 106 \\
Weighted RMS residuals ($\mu\text{s}$) & 21 & 11 & 18 & 755 \\
Reduced $\chi^2$ & 0.8 & 1.4 & 0.8 & 1.0 \\
\\
Galactic longitude, $l$ ($^\circ$) & 326.35 & 325.08 & 335.82 & 357.04 \\
Galactic latitude, $b$ ($^\circ$) & 1.94 & $-$2.05 & 2.79 & 2.92 \\
DM distance, $d$ (kpc) \\
\multicolumn{1}{c}{NE2001} & 2.9 & 1.9 & 2.4 & 3.8 \\
\multicolumn{1}{c}{YMW16} & 3.1 & 2.7 & 3.0 & 5.4 \\
Characteristic age, $\tau_\text{c}$ (Myr) & 6900 & 9100 & 30200 & 1400 \\
Surface magnetic field, $B_\text{surf}$ ($10^{10}\,\text{G}$) & 0.0331 & 0.0179 & 0.0136 & 0.30 \\
Spin-down luminosity, $\dot{E}$ ($10^{30}\,\text{erg}\,\text{s}^{-1}$) & 1880 & 3700 & 590 & 500 \\
\hline
  \multicolumn{4}{l}{$^{\text{a}}$ $m_\text{c,min}$ is calculated for an orbital inclination of $i = 90^\circ$ and an assumed pulsar mass of $1.4\,\text{M}_\odot$.}\\
  \multicolumn{4}{l}{$^{\text{b}}$ $m_\text{c,med}$ is calculated for an orbital inclination of $i = 60^\circ$ and an assumed pulsar mass of $1.4\,\text{M}_\odot$.}\\
\end{tabular}
\end{center}
\end{table*}

Work conducted over several decades \citep[e.g.][]{ts99,tv14} has implied the existence of a significant correlation between $P_\text{b}$ and $m_\text{c}$ for He-WD binaries, which is independent of the original low-mass progenitor star of the WD. With reference to Equation 20 of \cite{ts99}, this correlation can be used to estimate the mass of the He-WD companions of both pulsars, although this estimate should be treated with a degree of caution given the limited information available. Taking an average between the Pop. I and Pop. II cases outlined in \cite{ts99}, the estimated He-WD masses are $m_\text{c,calc}\simeq0.23\,\text{M}_\odot$ in the case of both pulsars.

Using these calculated WD masses (and assuming the mass of each pulsar), we can in turn estimate the inclination angles of the orbits of PSRs~J1537$-$5312 and J1547$-$5709 (with reference to Equation~\ref{eqn: mass function}). Assuming a canonical pulsar mass of $m_\text{p}=1.4\,\text{M}_\odot$ gives an inclination angle $i\simeq32^\circ$ for PSR~J1537$-$5312 and $i\simeq53^\circ$ for PSR~J1547$-$5709. Varying the pulsar mass between $1.3-2.0\,\text{M}_\odot$ allows the inclination angle to vary between $i=30^\circ-41^\circ$ for PSR~J1537$-$5312 and between $i=49^\circ-78^\circ$ for PSR~J1547$-$5709.  Naturally, the true value of $i$ in each case is highly dependent on the true mass of the pulsar, but it would appear from this analysis that the orbit of PSR~J1537$-$5312 is unlikely to be highly inclined, while the orbit of PSR~J1547$-$5709 may possess a high inclination should the pulsar be among the most massive of those currently known. A future detection of Shapiro delay in either pulsar would be able to provide an observational handle on the inclination angles and masses of both the pulsar and its He-WD companion, but such a detection will require more sensitive observations than are currently available from Parkes.

\subsection{PSR~J1618--4624, an unusual CO-WD binary MSP}

At first glance the $5.93$-ms pulsar PSR~J1618$-$4624 appears similar to PSRs~J1537$-$5312 and J1547$-$5709, as it is also a highly-recycled MSP in a circular orbit around a probable WD companion (see Table~\ref{tab: MSP-WD binary params}). However, with a minimum mass of $m_\text{c}=0.587\,\text{M}_\odot$, PSR~J1618$-$4624's companion is most likely a carbon-oxygen white dwarf (CO-WD), representing a much rarer class of MSP-WD binary which likely evolved from an intermediate-mass X-ray binary (IMXB) \citep{tauris11}. Only a handful of these systems are known to exist, including PSR~J1614$-$2230 \citep{dpr+10, tlk11, lrp+11, tlk12}, PSR~J1101$-$6424 \citep{ncb15} and recently PSR~J1933$-$6211 \citep{gvo+17}. No known counterpart to the CO-WD companion can be found in \textsc{SIMBAD} to within a radius of 30 arcseconds, and no further confirmation observations have been attempted.

PSR~J1618$-$4624 further stands out as a result of the unusual puzzle presented by the scenario of its formation and evolution. As described in \cite{tauris11}, the fully-recycled nature of this pulsar combined with a CO-WD companion favours a formation scenario involving Case~A RLO, in which a long and stable mass transfer occurs while the donor companion star is still on the main sequence (MS), thereby allowing the pulsar to fully spin-up. However, a Case~A scenario can only account for the $\sim1.78$-d orbit of PSR~J1618$-$4624 if the final mass of the neutron star is significantly lighter than the canonical value \citep[see e.g.][]{tlk11}, with `conventional' Case~A RLO scenarios favouring orbits $P_\text{b}\gtrsim3\,\text{d}$ \citep{tauris11}. Alternatively, the loss of angular momentum required to produce such a short orbit could imply a common-envelope (CE) evolutionary stage, during which the accreting NS becomes enveloped within the expanded outer layers of the donor star, a Case~C RLO scenario. However, a Case~C scenario cannot easily account for the observed degree of recycling \citep{tauris11}. 

This apparent conflict between the Case~A and Case~C RLO scenarios could potentially be solved by an eventual mass measurement of PSR~J1618$-$4624, which given the system's binary properties would most likely have to be derived from a measurement of the Shapiro delay. However, given the unknown inclination of the PSR~J1618$-$4624, the prospects of such a measurement from future observations remain uncertain. Even in the case of a fully-inclined orbit of $i=90^\circ$, the expected magnitude of the Shapiro delay is comparable to the current timing RMS of $\sim18\,\mu\text{s}$, with the individual TOA errors also being of a similar magnitude. More sensitive observations capable of reducing these effects will therefore be required to further investigate any potentially measurable Shapiro delay in this pulsar.

\subsection{PSR~J1745--23, a black widow pulsar}\label{subsec: black widow}

PSR~J1745$-$23 is a $5.42$-ms MSP in a black-widow class binary system. Follow-up observations with Parkes covering a 153-day span between MJD~57798 and MJD~57951 have allowed the determination of an approximate orbital solution which is sufficiently accurate so as to be useful as a folding model at all available epochs (see Table~\ref{tab: unsolved binary params}). The orbit appears to be highly circular, although in the absence of a phase-connected solution, we are presently unable to accurately constrain either the eccentricity $e$ or the longitude of periastron $\omega$. Therefore, these values are both fixed at $0$ in our model. Assuming a canonical pulsar mass of $m_\text{p}=1.4\,\text{M}_\odot$, PSR~J1745$-$23 would appear to have a rather light companion, with a minimum companion mass of $m_\text{c,min}\simeq0.027\,\text{M}_\odot$ and a median mass of $m_\text{c,med}\simeq0.031\,\text{M}_\odot$.

\begin{table}
\caption{Approximate binary parameters for PSR~J1745$-$23. Values in parentheses, where available, represent 1-$\sigma$ uncertainties on the final digit after weighting the TOAs such that $\chi^{2}=1$.}\label{tab: unsolved binary params}
\begin{center}
 \begin{tabular}{lc}
  \hline
  PSR name & J1745$-$23 \\
  \hline
  Fitting program & \textsc{tempo} \\
  Binary model & BT \\
  \\
  Orbital period, $P_\text{b}$ (d) & 0.165562(10) \\
  Projected semi-major axis, $x$ (lt-s) & 0.06247(6) \\
  Eccentricity, $e$ & 0$^\text{a}$ \\
  Longitude of periastron, $\omega$ ($^\circ$) & 0$^\text{a}$ \\ 
  Epoch of periastron, $T_0$ (MJD) & 57950.47559(3) \\
  \\
  Mass function, $f$ ($10^{-6}\,\text{M}_\odot$) & 9.550(16) \\
  Minimum companion mass$^{\text{b}}$, $m_\text{c,min}$ ($\text{M}_\odot$) & 0.02689(2) \\ 
  Median companion mass$^{\text{b}}$, $m_\text{c,med}$ ($\text{M}_\odot$) & 0.03111(4) \\
  \hline
  \multicolumn{2}{l}{$^{\text{a}}$ Values fixed at 0 due to lack of constraint and an}\\
  \multicolumn{2}{l}{evidently highly-circular orbit.}\\
  \multicolumn{2}{l}{$^{\text{b}}$ $m_\text{c,min}$, $m_\text{c,med}$ calculated per the assumptions in Table~\ref{tab: MSP-WD binary params}.}\\
 \end{tabular}
\end{center}
\end{table}

PSR~J1745$-$23 also appears to eclipse during superior conjunction. The duration of this eclipse is not yet precisely constrained, however a conservative upper limit of the eclipse duration is $T_\text{eclipse}<58\,\text{min}$, occurring between orbital phases of $0.14 < \varphi < 0.38$ (as specified by the model in Table~\ref{tab: unsolved binary params}).

We classify PSR~J1745$-$23 as a black widow (BW) based upon its lightweight companion, the presence of eclipses and the short orbital period ($P_\text{b} < 24\,\text{hr}$) \citep[see e.g.,][]{roberts13}. This may also be partly responsible for the difficulties that have been encountered while attempting to develop a phase-connected solution for this pulsar, a problem noted in other BW systems, e.g. PSR~J2051$-$0827 \citep{lvt+11, svf+16}. At present, although the companion can be identified as both lightweight and likely degenerate, any further classification remains uncertain. Ongoing observations will aim to establish a phase-coherent solution for this pulsar.

\subsection{PSR~J1727--2951, an intermediate-period, low-mass binary pulsar}

As can be seen in Figure~\ref{fig: discovery profiles}, PSR~J1727$-$2951 immediately stands out due to its unusually wide pulse profile. With $W_{50}=11.4$\,\text{ms} and $W_{10}=18.9$\,\text{ms}, PSR~J1727$-$2951 shows emission across the majority of its $28.4$-ms pulse profile. The measured $\dot{P}=3.1(6)\times10^{-19}$ indicates a partially recycled pulsar, which is consistent with its binary nature.

PSR~J1727$-$2951 is in a near-circular $\sim9.5$-hr orbit (Table~\ref{tab: MSP-WD binary params}). Based upon the system's mass function and an assumed pulsar mass of $m_\text{p}=1.4\,M_\odot$, the minimum mass for the companion is only $m_\text{c}\simeq0.014\,M_\odot$. Even taking an inclination angle $i=26^\circ$, which constitutes a 90\,\% upper limit on the mass of the companion \citep{lk05}, $m_\text{c}$ only increases to $\sim0.032\,M_\odot$, still well below a typical He-WD mass. The system would therefore appear to be superficially similar to the BW pulsar PSR~J1745$-$23 (see Section~\ref{subsec: black widow}), in both its companion mass and small projected semi-major axis. However, unlike a typical BW system, PSR~J1727$-$2951 is neither fully recycled nor does it display any evidence of eclipses, despite observational coverage of its entire orbital period. Nor does the pulsar appear to show the timing irregularities often typical of BW systems.

The true nature of PSR~J1727$-$2951 therefore remains puzzling. Should the orbit be inclined in such a way so as to prevent the observation of eclipses, a BW classification remains a possibility \citep{freire05}, as in the case of PSR~J2214$+$3000 \citep{rrc+11}. An alternative (although not mutually exclusive) view comes from the example of PSR~J1502$-$6752 \citep{kjb+12}, a pulsar with which PSR~J1727$-$2951 shares very similar values of both $P$ and $\dot{P}$. \cite{kjb+12} identify PSR~J1502$-$6752 as a member of the `very low mass binary pulsars' (VLMBPs), first described by \cite{fck+03}. It would appear that based on its low value of $m_\text{c}$, PSR~J1727$-$2951 also falls into this category, and may help bridge the gap between PSR~J1502$-$6752 and the rest of the VLMBP population. 

\subsection{PSR~J1653--45, a binary system with a long orbital period}

\begin{figure*}
\begin{center}
\includegraphics[height=\textwidth, angle=270]{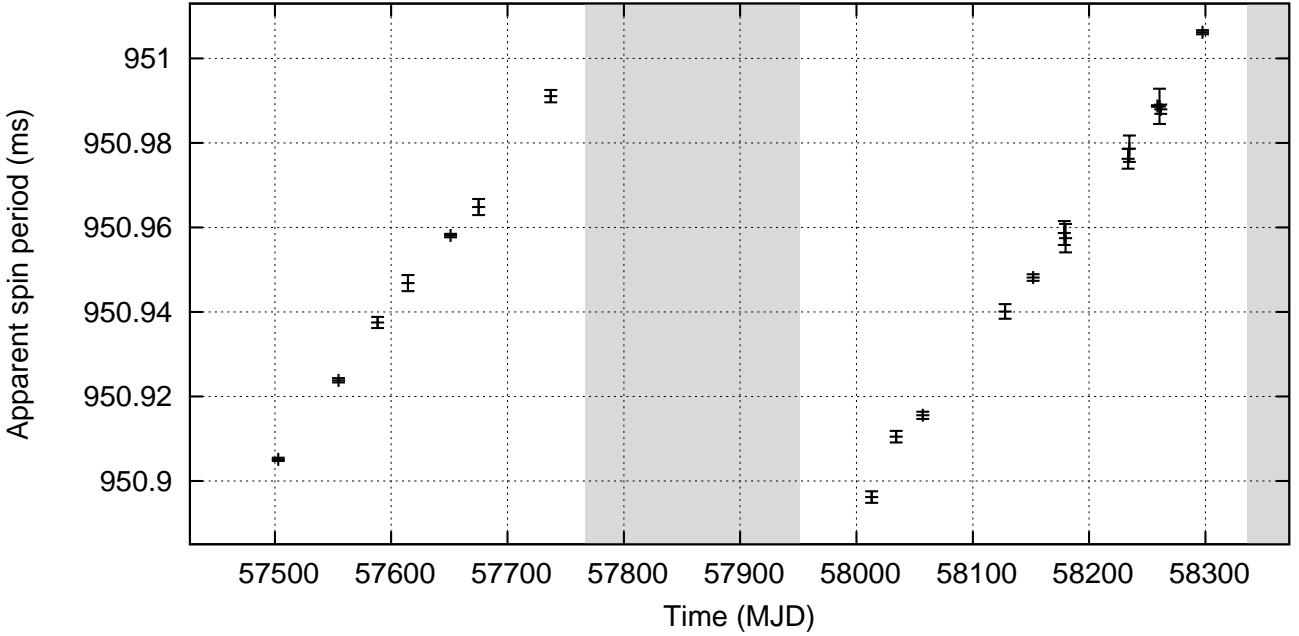}
\end{center}
\caption{Changes in the apparent spin period of PSR~J1653$-$45 over a span of approximately 2.2\,yr. Error bars represent the 1-$\sigma$ error on the measured period at each epoch. Grey regions indicate spans of time during which the pulsar appeared to be undetectable.}\label{fig: J1653-45}
\end{figure*}

PSR~J1653$-$45 is a $0.95$-s pulsar in a $\sim1.45$-yr orbit around an unknown binary companion which appears to eclipse the pulsar. Changes in the apparent spin period of PSR~J1653$-$45 can be seen in Figure~\ref{fig: J1653-45}. The cyclic nature of the pulsar's apparent spin period is clear evidence of a binary system, while the apparently predictable spans of non-detections as a function of orbital phase lends strong evidence to the eclipsing-binary hypothesis. When detectable, the apparent spin-period of the pulsar increases almost linearly, at an average rate of $\dot{P}\simeq 4.2\times10^{-12}$. From the two orbits we have observed thus far, we estimate an orbital period of $P_\text{b}\simeq528\,\text{d}$.

Due to the fact that the pulsar has remained undetected during roughly half of its orbital period, it has so far been impossible to develop either an orbital solution or a phase-connected timing solution. The near-linearity of its changing apparent spin-period during each orbit suggests that the system has a significant eccentricity, and estimates of the projected semi-major axis ($\sim500$ to $700\,\text{lt-s}$) indicate a companion mass on the order of a solar mass. Combined with the observed eclipses, we speculate that the companion of PSR~J1653$-$45 may be a MS star. If this is the case, PSR~J1653$-$45 may be similar to another pulsar discovered earlier in this survey, PSR~J1759$-$24 \citep{ncb15}, which is also considered likely to be in an eclipsing binary system with a long orbital period. Other potential examples of similar systems include PSR~B1820$-$11, a $\sim280$-ms pulsar in an eccentric $\sim1$-yr orbit around a companion speculated to either be a MS companion \citep{pv91} or a possible NS/WD companion \citep{tc99}, and PSR~B1259$-$63, a $\sim48$-ms pulsar in an eccentric $\sim3.4$-yr orbit around a $\sim10\,\text{M}_\odot$ Be-star companion and which also experiences a $\sim40$-day eclipse \citep{jml92}.

\subsection{PSR~J1706--4434, a glitching pulsar}

\begin{figure}
 \includegraphics[height=\linewidth, angle=270]{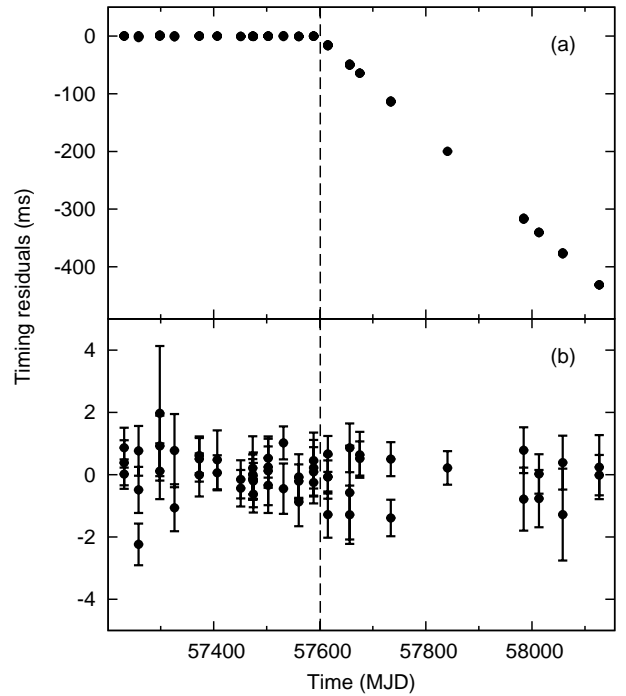}
\caption{Effect of the glitch observed in PSR J1706$-$4434 on the pulsar's timing residuals. Plot (a) shows the residuals plotted using the best pre-glitch phase-connected solution for the pulsar, without accounting for the glitch. Plot (b) shows the residuals plotted using the final phase-connected solution, which incorporates the glitch parameters listed in Table~\ref{tab: J1706 glitch parameters}. The glitch epoch (MJD 57601) is indicated by the dashed line.}\label{fig: J1706 glitch residuals}
\end{figure}

Following a confirmation and gridding observation on MJD~57169, timing observations with an approximately monthly cadence between MJD~57231 and MJD~57588 (a span of 357 days) were sufficient to develop a fully phase-connected solution for PSR~J1706$-$4434. However, all further timing observations (spanning MJD~57615 to MJD~58126) showed a clear deviation from this model, indicating a sudden decrease in the spin period of the pulsar. From this, it can be concluded that a glitch occurred in PSR~J1706$-$4434 at some point between MJD~57588 and MJD~57615. The effect of this glitch on the timing residuals of PSR~J1706$-$4432 is shown in Figure~\ref{fig: J1706 glitch residuals}.

Table~\ref{tab: J1706 glitch parameters} lists the parameters of the observed glitch in PSR~J1706$-$4434. Due to the limited cadence of timing observations surrounding the glitch, the precise glitch epoch cannot be accurately determined. A glitch epoch of MJD~57601, the approximate midway point between the epochs of the neighbouring observations, has therefore been assumed. Similarly, no evidence of a post-glitch relaxation has been detected, likely also due to the limited observational cadence as well as the limited timing precision of this pulsar. In addition, the permanent change to the spin-frequency derivative $\dot{\nu}$ (and corresponding change in the spin-period derivative $\dot{P}$) remains poorly constrained. Therefore, we present only an upper limit on this value in Table~\ref{tab: J1706 glitch parameters}, which represents an uncertainty of $3\,\sigma$.

\begin{table*}
 \caption{{Parameters of the glitch observed in PSR~J1706$-$4434. Values in parentheses represent standard 1-$\sigma$ uncertainties on the final digit as determined by \textsc{tempo2}. In the case of the permanent spin frequency derivative increment $\Delta\dot{\nu}$, an uncertainty of $3\,\sigma$ is provided as the upper limit.}}\label{tab: J1706 glitch parameters}

\begin{center}
 \begin{tabular}{lc}
  \hline
  \multicolumn{2}{c}{Glitch parameters for PSR~J1706$-$4434} \\
  \hline
  Estimated glitch epoch (MJD) & 57601 \\
  Phase increment, $\Delta\phi$ & 0.085(13) \\
  Permanent spin-frequency increment, $\Delta \nu$ (Hz) & $2.21(2)\times10^{-8}$ \\
  Permanent spin-frequency derivative increment, $\left|\Delta\dot{\nu}\right|$ (s$^{-2}$) & $<3.25\times10^{-17}$ \\
  \hline
 \end{tabular}
 \end{center}
\end{table*}

\subsection{PSR~J1812--15 and PSR~J1831--04, a pair of nulling, accelerated pulsars with long rotational periods}\label{subsec: long P0 binaries}

\begin{figure*}
\begin{center}
\includegraphics[height=0.49\textwidth, angle=270]{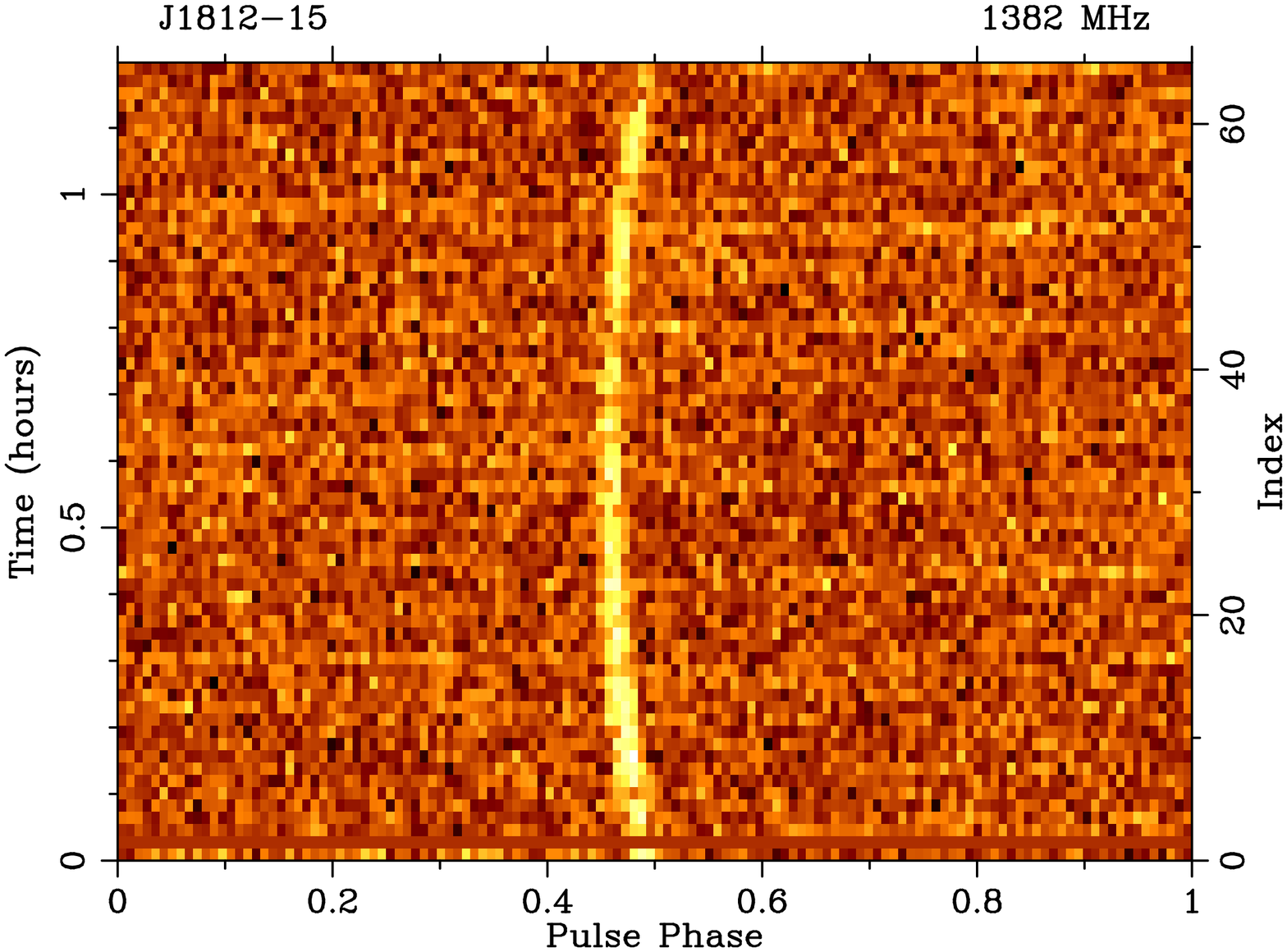}
\includegraphics[height=0.49\textwidth, angle=270]{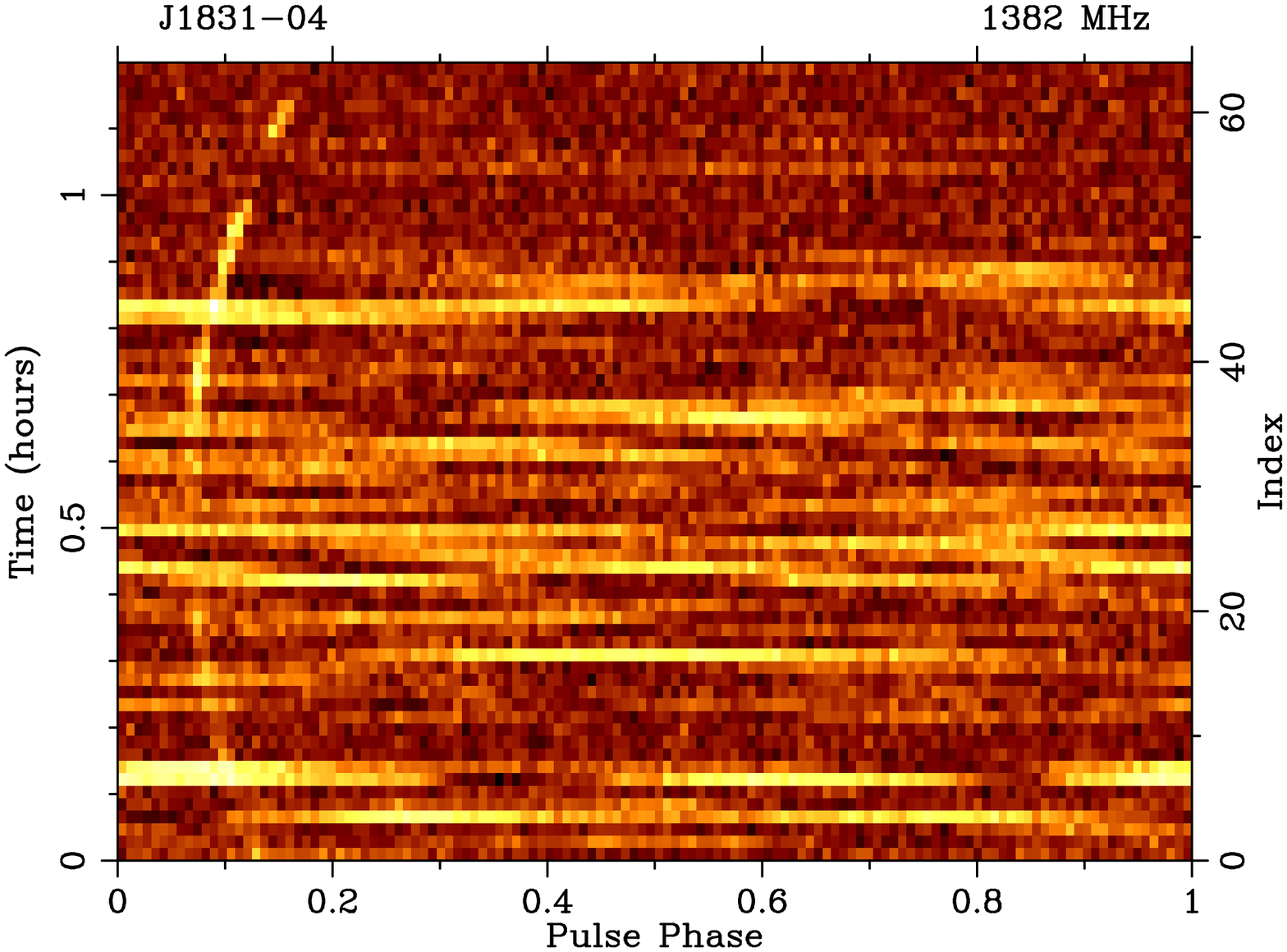}
\end{center}
\caption{The discovery observations of PSRs~J1812$-$15 (left) and J1831$-$04 (right). In both cases, these observations have been folded at a constant spin period and with an acceleration of $a=0\,\text{m\,s}^{-2}$. Even with the strong RFI present in the observation of PSR~J1831$-$04, both pulsars show clear evidence of acceleration (indicated by the parabolic curvature of each pulsar's phase over time), suggesting the presence of binary motion. There is some evidence for a brief null in PSR~J1831$-$04 approximately 1\,hr into the observation.}\label{fig: long-P binary accel}
\end{figure*}

PSR~J1812$-$15 and PSR~J1831$-$04 have rotational periods of approximately $1014$ and $1066\,$ms respectively. Evidence for acceleration was found in the discovery observations of both pulsars ($3.98(13)$ and $11.7(7)\,\text{m\,s}^{-2}$ respectively), indicating that they are likely to be in binary systems. The case for the binary nature of PSR~J1812$-$15 has been bolstered by a number of subsequent observations, during which its acceleration has been seen to vary between $3.62(11)$ and $5.57(9)\,\text{m\,s}^{-2}$ and its period between $1014.137(7)$ and $1014.93102(16)\,\text{ms}$, such that its binary nature is considered to be confirmed, although a complete orbital solution is currently unavailable. Meanwhile, despite a total of $\sim3.86\,\text{h}$ of follow-up observations between MJD~57798 and 58152, PSR~J1831$-$04 has yet to be redetected. The folded discovery observations of PSRs~J1812$-$15 and J1831$-$04 showing the acceleration in each pulsar can be found in Figure~\ref{fig: long-P binary accel}.

Efforts to determine the full orbital solutions of both PSR~J1812$-$15 and PSR~J1831$-$04 have been hindered by the fact that in addition to being in binary systems, both pulsars also appear to exhibit nulling behaviour, with changes between `on' and `off' states having been observed in both pulsars. The lack of any detection of PSR~J1831$-$04 since its initial discovery suggests that it may belong in the class of so-called `intermittent' pulsars whose `off' timescales may range from hours to days and even years, and of which only a handful of examples are currently known \citep[see e.g.][]{klob+06, crc+12, llm+12, lsf+17}. As our understanding of the nulling timescales of PSR~J1812$-$15 still remains incomplete, it is possible that this pulsar may also qualify as intermittent, with non-detections sometimes appearing to last longer than the pulsar's typical integration time ($\sim10-30$\,min).

A rudimentary estimate of the nulling fraction (NF) of each pulsar (the fraction of time each pulsar spends in its `off' state) was conducted by visually inspecting each folded observation, in a method analogous to that described by \cite{lsf+17}. Intervals during which the pulsar was unambiguously detectable were classified as `on', while all other intervals were classified as `off'. The provided error $\sigma$ on each NF is the standard error for a randomly sampled data set with $n$ samples,
\begin{equation}
\sigma = \sqrt{\frac{\text{NF}\left(1-\text{NF}\right)}{n}}.
\end{equation}
The calculated NF's for PSRs~J1812$-$15 and J1831$-$04 are given in Table~\ref{tab: nulling table}. It should be noted that all observations of PSR~J1812$-$15 after MJD~58013 were recorded in fold mode using a fixed spin period. Should the binary orbit of the pulsar be more extreme than anticipated, this may also account for a portion of the observed non-detections, lowering the pulsar's true NF. 

\begin{table*}
\caption{Parameters of 4 newly-discovered pulsars which display either nulling or intermittent behaviour. For each pulsar, the span over which suitable observations are available, the number of observations and the total integration time are provided. The nulling fraction (NF) and its uncertainty is calculated according to the description in Section~\ref{subsec: long P0 binaries}.}\label{tab: nulling table}
\begin{center}
\begin{tabular}{lcccc}
\hline
PSR name & Observing span (MJD) & Num. observations & Total integration time (hr) & NF(\%) \\
\hline
J1810$-$1709 & 55926$-$58407 & 5 & 3.5 & 70(20) \\
J1812$-$15 & 55845$-$58298 & 48 & 18.0 & 54(7) \\
J1831$-$04 & 55923$-$58152 & 18 & 4.8 & 79(10) \\
J1854$-$0524 & 56028$-$57609 & 10 & 4.2 & 72(14) \\
\hline
\end{tabular}
\end{center}
\end{table*}

\subsection{PSR~J1822--0902, a pulsar with significant timing noise}\label{subsec: J1822}

PSR~J1822$-$0902 is notable in that it possesses a high spin-down rate of approximately ${\dot{P}=1.77809(8)\times10^{-14}}$, implying both a young characteristic age $\tau_\text{c}$ and a high spin-down luminosity $\dot{E}$. In fact, PSR~J1822$-$0902's estimated characteristic age of $\tau_\text{c}=0.132\,\text{Myr}$ ranks as the smallest value of $\tau_\text{c}$ among the pulsars presented in this paper, while its $\dot{E}=2.13\times10^{35}\,\text{erg\,s}^{-1}$ ranks as the highest. In addition, PSR~J1822$-$0902 displays a significant degree of timing noise, requiring the use of additional spin derivative terms to model its observed behaviour. At present, attempts to model this timing noise as the result of either parallax (unlikely given the large DM-distance estimates listed in Table~\ref{tab: solved pulsars main}), proper motion or binary motion have been unsuccessful. It would therefore appear likely that this timing noise is intrinsic to the pulsar.

\subsection{PSR~J1810--1709, a nulling pulsar}

With a DM of $\sim670\,\,\text{cm}^{-3}\,\text{pc}$, PSR~J1810$-$1709 immediately stands out due to the presence of what may be a scattering tail in its profile, as seen in Figure~\ref{fig: discovery profiles}. Alternatively it may be that the pulsar simply has a wide inherent pulse profile, although some amount of scattering is likely given the high DM. Future multi-frequency or wide-band observations of this pulsar may be useful in investigating the role played by scattering in the profile of this pulsar.

One feature that is not immediately evident from the pulsar profile is the nulling behaviour exhibited by PSR~J1810$-$1709. Unlike the other variable pulsars we report in this paper, the `on' and `off' timescales of PSR~J1810$-$1709 appear to be much shorter. The average nulling and emission timescales are $333\,\text{s}$ and $135\,\text{s}$ respectively, with corresponding standard deviations of $291\,\text{s}$ and $122\,\text{s}$. All of these values are on the order of 100 to 300 pulse periods. With reference to the definition provided in Section~\ref{subsec: long P0 binaries}, this would appear to classify PSR~J1810$-$1709 as a nulling pulsar (as opposed to an intermittent pulsar). Once again, we calculate a NF (given in Table~\ref{tab: nulling table}) by adopting the same method as described in Section~\ref{subsec: long P0 binaries}, using the limited amount of search mode data recorded with the Parkes radio telescope.

\subsection{PSR~J1854--0524, a potentially intermittent pulsar}

The one feature of note regarding PSR~J1854$-$0524 is its apparent variability. Although approximately monthly confirmation observations commenced on MJD~57405, it was not until MJD~57554 that the pulsar was redetected, and not until MJD~57609 that an observation was recorded with sufficient S/N so as to be able to grid the pulsar's position. Using the same method as described in Section~\ref{subsec: long P0 binaries}, we are able to derive an estimate of the pulsar's NF as provided in Table~\ref{tab: nulling table}. As the longest non-detection of this pulsar exceeds 30\,min, it may be that PSR~J1854$-$0524 can also be classed among the intermittent pulsars, although further study of its `on' and `off' timescales will be required before a conclusive determination can be made.

\section{Comparison to the known pulsar population}\label{sec: pop comparison}

We round out this paper with an updated comparison of the discoveries of the HTRU-S LowLat pulsar survey to the known pulsar population within the survey region. We therefore consider the \PSRnumHTRUTOTAL pulsars reported both here and in \cite{ncb15}. For the \PSRnumADC pulsars reported in this paper, we use the parameters listed in Section~\ref{sec: new discoveries}, with the exception of PSR~J1757$-$1854, whose parameters are given in \cite{cck+18}. For the remaining \PSRnumNCB pulsars, we use the parameters listed in \cite{ncb15}, with the exception of PSR~J1755$-$2550, whose parameters are taken from \cite{nkt+18}. For the population of known pulsars, we have used the same set of pulsars within the survey region in each analysis, with their parameters taken from \textsc{psrcat}. However, as each known pulsar may not have a full set of available parameters, the number of known pulsars used will vary between analyses.

\begin{figure}
\begin{center}
\includegraphics[height=\linewidth, angle=270]{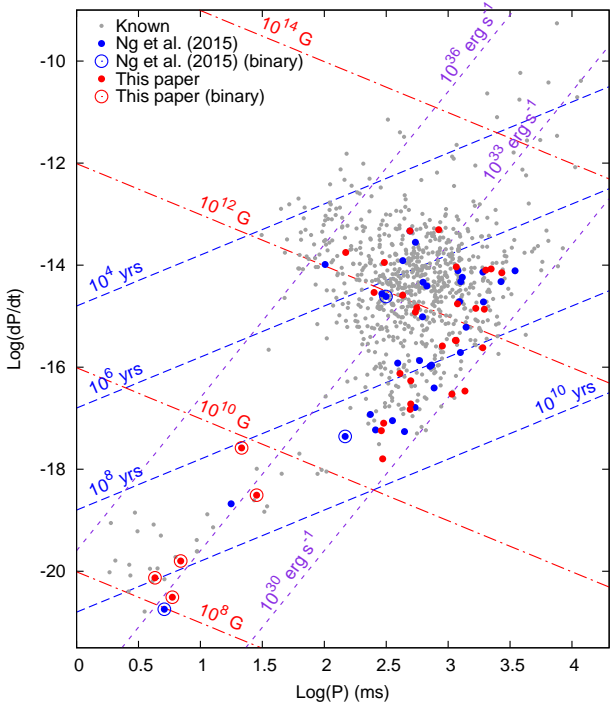} 
\end{center}
\caption{A $P$-$\dot{P}$ diagram displaying \PSRnumTOTALsolved HTRU-S LowLat discoveries with well-constrained $P$ and $\dot{P}$ against the known Galactic pulsar population within the survey area. The \PSRnumNCBsolved solved pulsars from \protect\cite{ncb15} are shown in blue, while the \PSRnumADCsolved solved pulsars from Tables~\ref{tab: solved pulsars main}, \ref{tab: solved pulsars supplementary} and \ref{tab: MSP-WD binary params} are shown in red. Grey points represent the population of \PSRnumKNOWNsurveyPPDOT known pulsars within the survey region. Circled points indicate binary pulsars.}\label{fig: discovery p-pdot} 
\end{figure}

For the \PSRnumTOTALsolved pulsars in the HTRU-S LowLat survey population with well-constrained values of $P$ and $\dot{P}$, Figure~\ref{fig: discovery p-pdot} shows their positions on a $P$-$\dot{P}$ diagram as compared to \PSRnumKNOWNsurveyPPDOT previously-known pulsars within the survey region. It would seem from Figure~\ref{fig: discovery p-pdot} that the HTRU-S LowLat pulsars represent a generally older, more-evolved collection of pulsars. This trend appears to be shared between the discoveries reported from both portions of the survey processing, however as just under half of the reported discoveries lack a well-measured $\dot{P}$, the possibility remains that this apparent trend may be the result of small-number statistics, a caveat which will apply throughout this section.

\subsection{Distance}\label{subsec: distance}

As described in Section~\ref{sec: new discoveries}, we estimate the distance to each of the \PSRnumHTRUTOTAL HTRU-S LowLat pulsars via their measured DM values, using independent estimates from both the NE2001 \citep{NE2001a} and YMW16 \citep{YMW16} electron density models. As per \cite{ncb15}, a typical uncertainty of $25\,\%$ is assumed for the NE2001 distance estimates, while \cite{YMW16} report a typical uncertainty of $\sim10\,\%$ for distance estimates from the YMW16 model. Note that it is not the intent of this comparison to render an assessment of the accuracy of either of these models, but in light of the statistical uncertainties involved in pulsar distance estimation, it is prudent to consider both of them.

Contrary to \cite{ncb15}, who reported the discovery of no pulsars within a distance less than $\sim2\,\text{kpc}$, we report the discovery of 2 such nearby pulsars. PSR~J1753$-$28, with a DM of only $18.0(9)\,\text{cm}^{-3}\,\text{pc}$ and a DM distance of $0.6-0.7\,\text{kpc}$, ranks as the closest pulsar discovered in the HTRU-S LowLat survey. Meanwhile PSR~J1514$-$5316, with a DM of only $27.1(3)\,\text{cm}^{-3}\,\text{pc}$ and a DM distance of $\sim0.9\,\text{kpc}$, comes in at a close second. This would indicate an approximate fraction of `nearby' pulsars discovered within the survey of $\sim2\,\%$ which, accounting for small-number statistics, is consistent with the $\sim3.8\,\%$ expectation value calculated by \cite{ncb15}.

\begin{figure*}
\begin{center}
\includegraphics[height=\linewidth, angle=270]{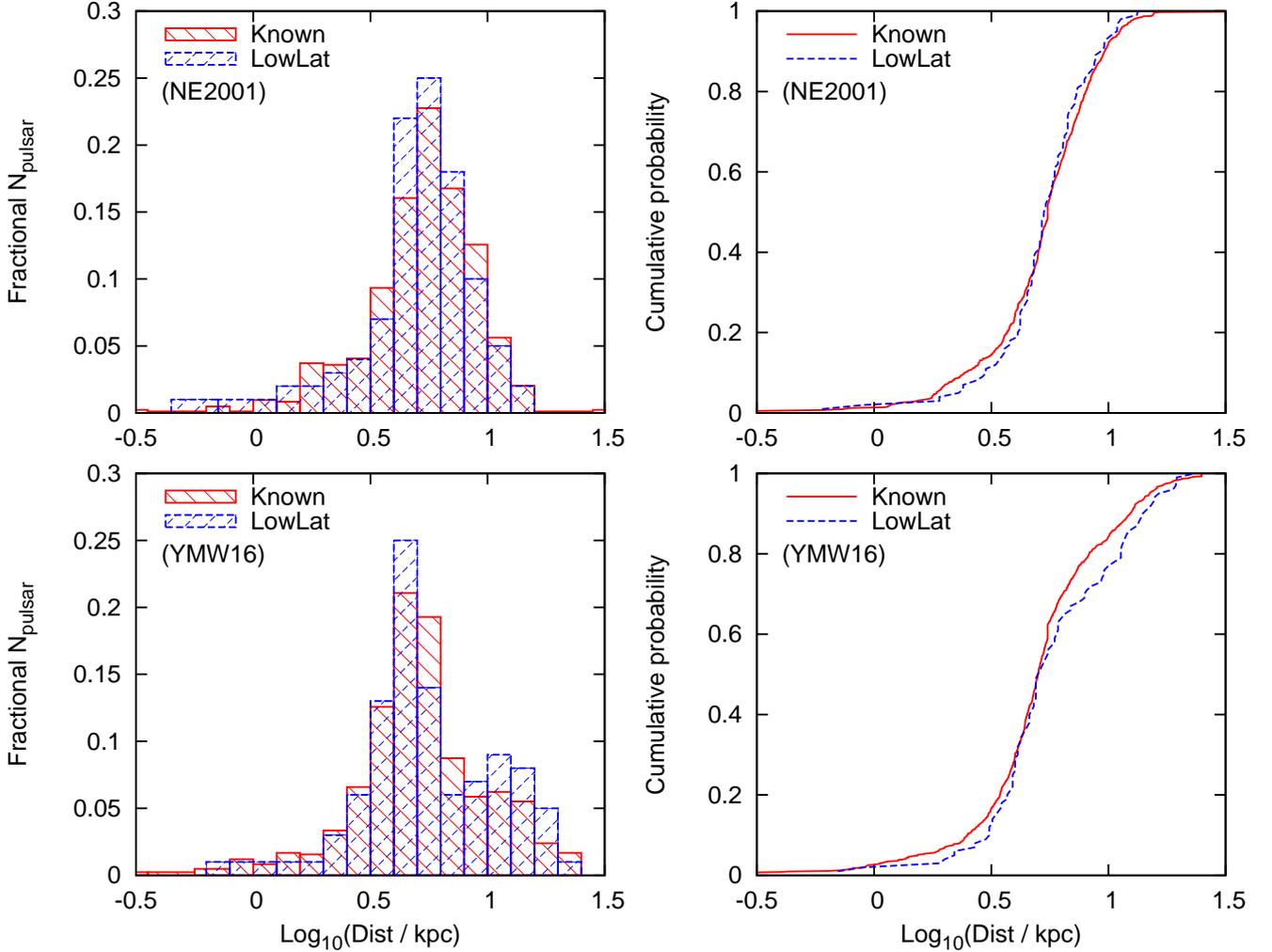}
\end{center}
\caption{Distance distributions of the \PSRnumHTRUTOTAL HTRU-S LowLat pulsars (blue) and \PSRnumKNOWNsurveyDISTANCE previously-known Galactic pulsars (red). Shown on the left are histograms of the distance distributions of each population under both the NE2001 (top) and YMW16 (bottom) models, normalised by the number of pulsars in each distribution and with a binsize of 0.1. Shown on the right are the corresponding CDFs of each distribution, again featuring the NE2001 (top) and YMW16 (bottom) models.}\label{fig: distance distribution} 
\end{figure*}

A more comprehensive assessment of the degree to which the distances of the HTRU-S LowLat pulsars correspond to those of the overall Galactic population can be made with a Kolmogorov-Smirnov (KS) test. Distance statistics of the known pulsars within the survey region were compiled from \textsc{psrcat}. Where precision distance measurements were unavailable from \textsc{psrcat}, DM-distance estimates were calculated once again using both the NE2001 and YMW16 models, giving two data sets consisting of distance estimates for \PSRnumKNOWNsurveyDISTANCE known pulsars. The distance statistics of the HTRU-S LowLat and known populations were then compared on a per-model basis.

In both cases, the null hypothesis (that the distribution of distances derived for the \PSRnumHTRUTOTAL HTRU-S LowLat discoveries is drawn from the same population distribution as the known pulsars) cannot be ruled out. The YMW16 test results in a $p$-value of $\sim0.3$, while the NE2001 test results in a $p$-value of $\sim0.6$, with the $p$-value indicating the probability of the null hypothesis being correct. It would therefore appear that the distances of the HTRU-S LowLat pulsars are consistent with the background population. This consistency is shown in Figure~\ref{fig: distance distribution}, which shows the similarities in the distance distributions of the two populations for both the NE2001 and YMW16 models both in terms of their respective histograms and cumulative distribution functions (CDFs).

\subsection{Luminosity}\label{subsec: luminosity}

\begin{figure*}
\begin{center}
 \includegraphics[height=\linewidth, angle=270]{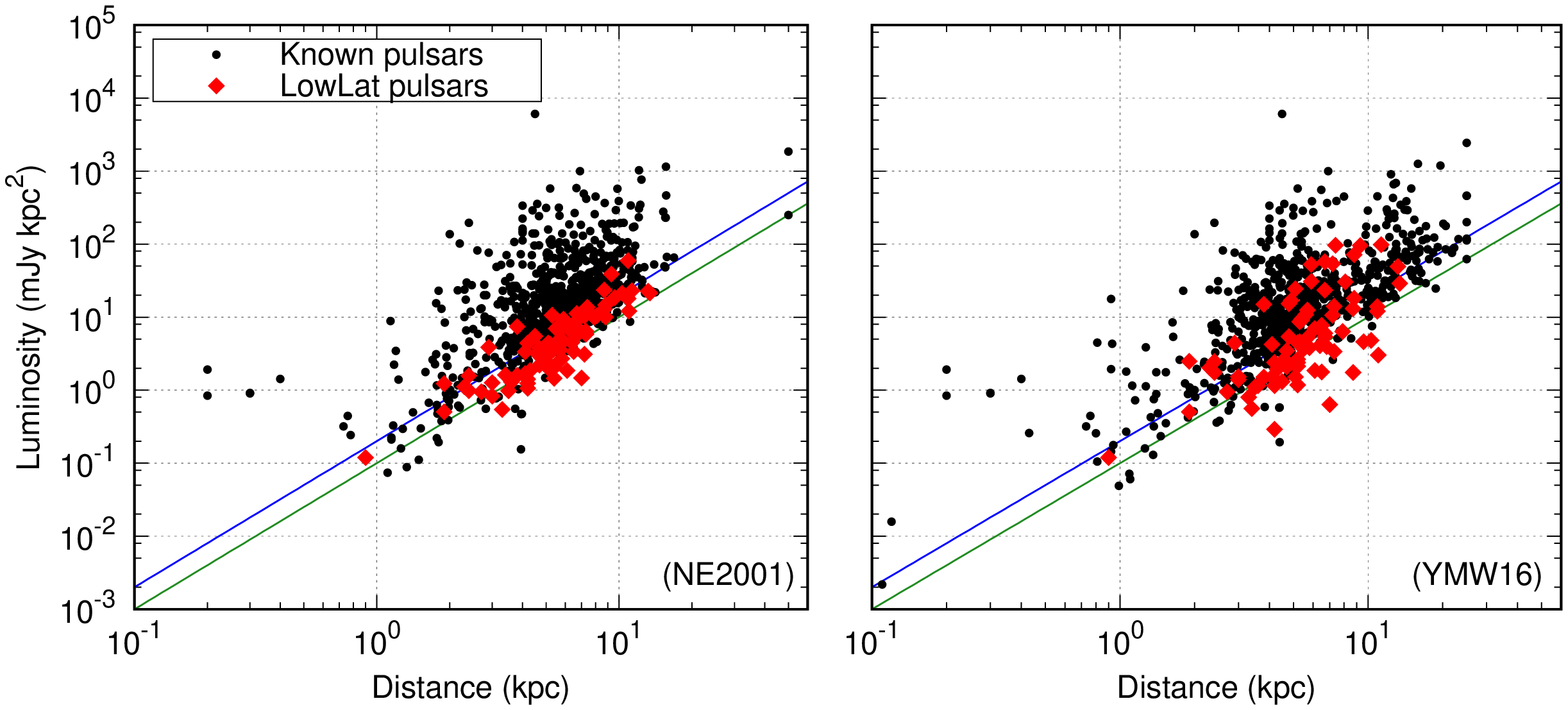} 
\end{center}
\caption{A comparison of the distances and 1.4-GHz luminosities of the \PSRnumHTRUFLUX HTRU-S LowLat pulsars with calibrated flux densities to the background pulsar population. HTRU-S LowLat pulsars are shown in red, while the \PSRnumKNOWNsurveyFLUX known pulsars within the survey region are shown in black. The green and blue lines represent lines of constant 1.4-GHz flux density at $0.1$ and $0.2\,\text{mJy}$ respectively. Where more accurate distances are unavailable, distances and luminosities are estimates using the NE2001 (left) and YMW16 (right) electron density models.}\label{fig: luminosity} 
\end{figure*}

At present, \PSRnumHTRUFLUX of the HTRU-S LowLat pulsars have calibrated flux density measurements at $1.4\,\text{GHz}$. Using the distance estimates described in Section~\ref{subsec: distance}, it is therefore possible to also conduct a population analysis based on luminosity. This analysis was conducted against a background population of \PSRnumKNOWNsurveyFLUX known pulsars within the survey area which also have measured $1.4\,\text{GHz}$ flux density values. Figure~\ref{fig: luminosity} shows a comparison of these two populations, with distances and luminosities determined using both the NE2001 and YMW16 models. Note that the luminosity values (as listed here in Table~\ref{tab: discovery summary}) should be treated with a degree of caution, as they rely on both the uncertainties of the measured flux density of each pulsar as well as its DM distance estimates (see Section~\ref{subsec: distance}).

From both the NE2001 and YMW16 luminosity estimates shown in Figure~\ref{fig: luminosity}, it can clearly be seen that the HTRU-S LowLat survey has met its objective of uncovering pulsars at the lower end of the luminosity distribution. This is verified by a KS test, which shows that for both the NE2001 and YMW16-based luminosity estimates, the probability of the HTRU-S LowLat luminosity distribution having being drawn from the background pulsar distribution is less than $0.0001\,\%$.

This result, although still somewhat preliminary, becomes more significant when considered together with the results of Section~\ref{subsec: distance}. That is, we have clearly discovered pulsars with lower luminosities across all distance scales, while sampling the same distance distribution as that of the background pulsars. This would seem to imply that we have yet to reach a `bottom' of the luminosity distribution, and that even lower-luminosity pulsars are likely to still exist. This appears to be true even in the case of nearby pulsars, whose numbers one might naively expect to be exhausted first. Were this not the case, we should expect to see a growing bias in our discovered population towards larger distances when compared against the background population. Bolstering this conclusion are the particular examples of PSR~J1514$-$5316, which has an estimated luminosity of only $\sim0.1\,\text{mJy}\,\text{kpc}^2$ under both the NE2001 and YMW16 models and ranks as the lowest-luminosity pulsar of the HTRU-S LowLat population by at least a factor of 2, and the binary MSP PSR~J2322$-$2650, a recent discovery of HTRU-S HiLat \citep{sbb+18} whose luminosity at $1.4\,\text{GHz}$ and DM distance estimates appear extremely similar to PSR~J1514$-$5316. It will likely fall to the next generation of radio telescopes such as MeerKAT, FAST and the SKA, to continue the exploration into just how low the pulsar luminosity distribution truly extends.

\subsection{Characteristic age}\label{subsec: characteristic age}

A further comparison can be performed using the characteristic age distribution of the HTRU-S LowLat pulsars. As per \cite{lk05}, the characteristic age $\tau_\text{c}$ is derived from measurements of $P$ and $\dot{P}$, therefore only those pulsars for which these values are both well measured are considered in this analysis. Similarly, as $\tau_\text{c}$ becomes contaminated as a result of the recycling process, only unrecycled pulsars (here defined as those pulsars for which the surface magnetic field $B_\text{surf}<3.0\times10^{10}\,\text{G}$ in order to remain consistent with the analysis performed by \cite{ncb15}) are considered. These requirements leave a total of \PSRnumHTRUAGE HTRU-S LowLat pulsars and a background population of \PSRnumKNOWNsurveyAGE known pulsars.

\begin{figure*}
\begin{center}
\includegraphics[height=\linewidth, angle=270]{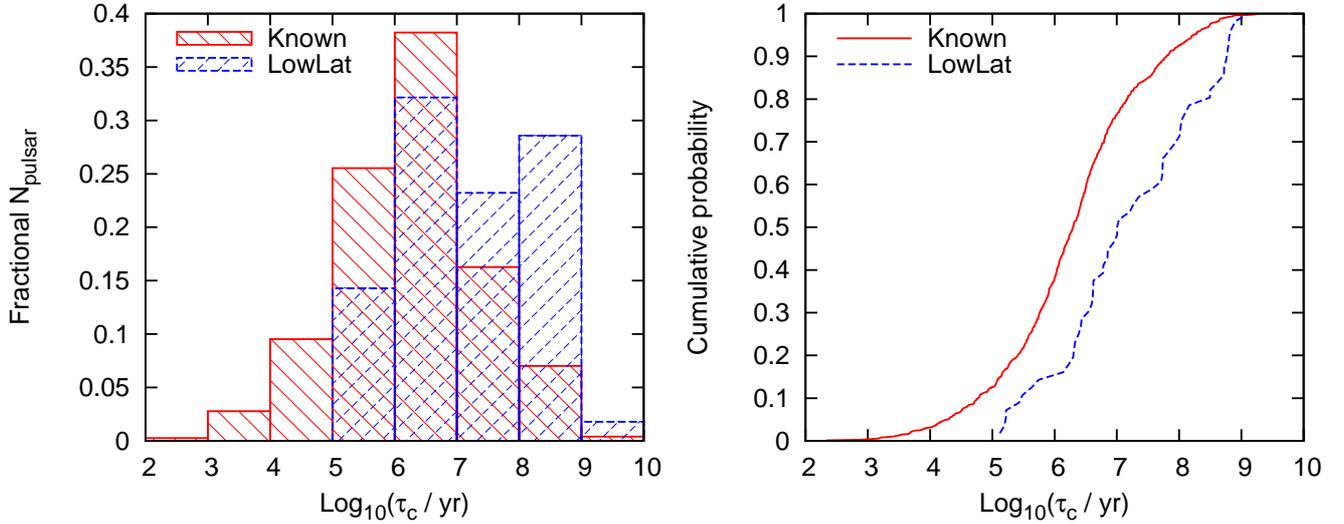}
\end{center}
\caption{Characteristic age ($\tau_\text{c}$) distributions of the \PSRnumHTRUAGE HTRU-S LowLat pulsars (blue) and \PSRnumKNOWNsurveyAGE previously-known Galactic pulsars (red) with well-constrained values of $P$ and $\dot{P}$. Shown on the left are the histograms of the $\tau_\text{c}$ distributions of each population, normalised by the number of pulsars in each distribution and with a binsize of 1.0. Shown on the right are the corresponding CDFs of each $\tau_\text{c}$ distribution.}\label{fig: age distribution} 
\end{figure*}

A KS test between these two distributions (as shown in Figure~\ref{fig: age distribution}) shows that they are not consistent with being drawn from the same population, to within a probability of $<0.001\,\%$. From both of these side-by-side comparisons, it is clear that the sample of HTRU-S LowLat pulsars does indeed represent an older population of pulsars, with no unrecycled pulsar yet having a $\tau_\text{c}\leq0.1\,\text{Myr}$ (a frequent definition of a `young' pulsar). This reinforces a trend in the HTRU-S LowLat pulsar population first noted by \cite{ncb15}, and stands in opposition to the earlier prediction of \cite{bates12} who, after similarly not discovering any young, unrecycled pulsars in the HTRU-S MedLat survey, predicted that such pulsars would likely be uncovered by HTRU-S LowLat due to its higher sensitivity. It is possible that this result may still be attributable to the ambiguities of small-number statistics, and continued timing of the remaining unrecycled and unsolved pulsars will assist in further clarifying this result.

\section{Survey yield evaluation}\label{sec: survey yield}

As noted previously, the processing of the HTRU-S LowLat survey now stands at $\sim94\,\%$. Therefore, it is now possible to conduct a near-full evaluation of the yield of both the survey and its first-pass processing pipeline with respect to both earlier predictions and the earlier analysis carried out by \cite{ncb15}. \cite{kjvs10} estimated that 957 `normal' pulsars (defined for this analysis as pulsars with $P>30\,\text{ms}$ in order to maintain consistency with earlier analyses) should be detected by the HTRU-S LowLat survey, while \cite{ncb15} later revised this estimate to 1020 normal pulsars. As these two results are statistically consistent, the latter is adopted to maintain consistency with the earlier yield analysis in \cite{ncb15}. Rescaling this estimate to account for the small proportion of the survey yet to be processed and reviewed results in a final estimate of $\sim960$ normal pulsars which ought to have been detected by current processing.

By comparison, as reported in Section~\ref{sec: redetections}, a total of 649 unique pulsars were redetected in the processed $\sim94\,\%$ of the HTRU-S LowLat survey, of which 631 are normal pulsars. This is in addition to the 58 normal pulsars reported by \cite{ncb15} and the 34 normal pulsars reported here, giving a total of 723 normal pulsar detections. Evidently, this falls short of the projected $\sim960$ detections by a factor of $\sim25\,\%$. This result also stands in contrast to the earlier evaluation of \cite{ncb15}, who predicted an expectation of $470-510$ normal pulsar detections from the initial survey processing and reported a consistent 485 normal pulsar detections. The apparently lower detection rate in the portion of the survey data processed in this paper is in part due to a common set of pulsars which were detected in both processed portions (as noted previously in Section~\ref{sec: redetections}), although this cannot fully account for the discrepancy between the predicted and actual number of detected normal pulsars.

Separate predictions were also produced for the MSP population (here defined as $P\leq30\,\text{ms}$ again in order to maintain consistency with earlier analyses), as these pulsars represent a distinct population which in turn is subject to different selection biases in a blind pulsar survey. \cite{kjvs10} and \cite{ncb15} respectively predicted 51 and 43 MSP detections for the HTRU-S LowLat survey region, while an in-depth study by \cite{lbb+13} predicted a higher MSP yield of 68. Taking the lower and upper limits of these estimates and rescaling to account for the $\sim94\,\%$ completion of the survey processing gives a final estimate of $40-64$ expected MSP detections. By comparison, only 18 of the redetected pulsars outlined in Section~\ref{sec: redetections} meet the $P \leq30\,\text{ms}$ criteria, along with 2 pulsars reported by \cite{ncb15} and 6 additional pulsars reported here. This gives a total of only 26 MSP detections from HTRU-S LowLat, a factor of $1.5-2.5$ lower than predicted.

Multiple potential reasons exist as to why our detections of both normal pulsars and MSPs have fallen below their initial predictions. With respect to the MSP population, \cite{ncb15} and \cite{ekl+13} have already put forward a number of potential causes, including a limited understanding of the underlying Galactic MSP population leading to inadequate model estimates, detectability limitations due to scatter broadening, and the inability of our segmented search technique to take advantage of the fully-coherent sensitivity of each $72$-min observation. While the discovery of PSR~J1757$-$1854 shows the strengths of our segmented search pipeline in being able to detect pulsars with short, highly-accelerated orbits, other similar pulsars such as PSRs~J1802$-$2124 and J1141$-$6545 may still have been missed for the reasons discussed in Section~\ref{subsec: binary redetections}. Therefore, the implementation of a fully-coherent binary search remains an ongoing goal of the future reprocessing of this survey.

Additional factors exist which are likely to influence the lack of detections of both normal pulsars and MSPs. Nulling and intermittency, often factors in longer-period pulsars \citep[see e.g.][]{biggs92}, remain unavoidable obstacles in blind pulsar surveys. That being said, the $72$-min integration times of the HTRU-S LowLat survey have been able to partially mitigate this effect, with at least six either nulling or intermittent pulsars and at least three eclipsing pulsars having been discovered so far. The influence of RFI also remains an ongoing problem, as despite our use of multiple RFI mitigation techniques, it accounted for a significant fraction of the candidates reported by the search pipeline. In particular, certain beams of the MB20 receiver would often experience extended spans of worsened interference, indicating potential hardware problems which multibeam RFI excision techniques would be unable to correct for. However, while RFI is likely to have contributed to a degraded survey sensitivity, its influence remains difficult to quantify.

Another limitation lies in the inherent shortcomings of Fourier-based search techniques.  For example, the increase in Fourier power caused by red noise contamination (a consequence of our long integration times) is likely to mask the presence of longer-period pulsars in a Fourier-based search, as highlighted by \cite{lbh15}. Fourier searches also lose sensitivity due the fact that they are an incoherent search technique, and are only able to sum a limited number of harmonics, resulting in the loss of additional Fourier power. This is especially true in the case of long-period or narrow duty-cycle pulsars, for which additional S/N can be attained once the candidate is folded at the correct period. \cite{lbh15} further demonstrated that in the case of the Pulsar Arecibo L-band Feed Array survey \citep[PALFA;][]{cfl+06}, a significant loss of sensitivity was experienced for pulsars with $P>100\,\text{ms}$, becoming especially noticeable for pulsars with $P>1\,\text{s}$, with approximately $35\,\%$ of simulated pulsars remaining undetected. The synthetic profile injection technique used by \cite{lbh15} in evaluating this loss in sensitivity may be useful in more thoroughly evaluating  the limitations in the Fourier transform search technique specific to the HTRU-S LowLat data set. However, such an analysis lies beyond the scope of this paper, and may be included in a future publication.

One technique which may be able to counter these limitations of Fourier-based search techniques is that of the `Fast Folding Algorithm' \citep[FFA,][]{sta69}. Following recent in-depth studies of the FFA \citep[see e.g.][]{kml09, cbc+17}, an implementation of the FFA is now being used as part of a new processing pass of the HTRU-S LowLat survey. Additionally, folding of candidates with lower Fourier significance may also reveal new candidates whose detectability was otherwise affected by the limitations of incoherent harmonic summing. Both of these techniques have already already produced a handful of additional pulsar discoveries which will be the subject of a future publication. It is too soon to know to what extent these new discoveries will be able to bridge the gap between the expected and actual numbers of pulsars both detected and newly-discovered in the HTRU-S LowLat survey. However, it is clear that the HTRU-S LowLat survey will remain a valuable resource for the application of future algorithms and for the verification of pulsars detected in future surveys.

\section{Additional Discussion \& Conclusions}\label{sec: conclusions}

The HTRU-S LowLat pulsar survey was undertaken with the twin primary goals of discovering short-$P_\text{b}$, relativistic binary pulsars, as well as pulsars situated at the low end of the luminosity distribution. From the results presented both here and in \cite{ncb15}, which span the discovery of \PSRnumHTRUTOTAL new pulsars, it is clear that the survey has been successful on both counts. While PSR~J1757$-$1854 is the only relativistic binary to have been discovered by the survey, as many as 12 other binary systems have also been discovered, verifying our sensitivity to binary systems. Meanwhile Section~\ref{subsec: luminosity} clearly indicates that the discovered pulsars sit lower on the luminosity distribution function than the general background population within the survey region, and suggests that there are pulsars with even lower luminosities yet to be discovered. These achievements come as a direct consequence of the longer integration times, larger bandwidth and finer time resolution employed by this survey in comparison to previous search efforts such as those undertaken with the PMPS, as well as our implementation of the `partially-coherent segmented acceleration search' technique.

However, it is clear that there remains room for improvement. For example, although our pipeline only missed approximately $2\,\%$ of the known pulsar population, the examples of PSRs~J1435$-$6100 and J1810$-$2124 in Section~\ref{subsec: binary redetections} and the lower than anticipated survey yield show that there remains parameter spaces to which the pipeline is insensitive. Therefore, in addition to completing the processing of the entire survey with the segmented pipeline to ensure a consistent first-pass processing, future processing should focus on resolving these known shortcomings. This should include the development and implementation of a fully-coherent binary search (e.g. a `jerk' search), as well as an implementation of the FFA. With these and other modifications, it is likely that reprocessing of the HTRU-S LowLat survey will continue to yield new results for many years to come. Even with the ongoing Survey for Pulsars and Extragalactic Radio Bursts \citep[SUPERB;][]{kbj+18} and the advent of the next generation of pulsar surveys which will be enabled by telescopes such as MeerKAT, FAST and the SKA, the steady production of new pulsar discoveries and other key scientific results from the multiple reprocessings of the PMPS \citep[see e.g.][]{kle+10,emk+10,ekl+13} clearly demonstrates the value in reprocessing legacy surveys with a variety of search techniques.

In addition to having met its broad scientific goals, we have also reported the discovery of a number of scientifically-interesting pulsars from the HTRU-S LowLat survey, many of which are likely to require continuing observation in order to fully exploit their scientific potential. For example, the three MSP-WD binary systems (PSRs~J1537$-$5312, J1547$-$5709 and J1618$-$4624) add to the collective picture of MSP evolution, with the latter of these presenting lingering questions with regards to its evolutionary history and whether it should be classed as the result of either Class A or Class C RLO. Ongoing observations of PSR~J1618$-$4624 (especially with an increase in sensitivity) may lead to an eventual Shapiro measurement of the system should its inclination be sufficiently high, helping to resolve this ambiguity via a measurement of the pulsar mass. Meanwhile, the observed glitch in PSR~J1706$-$4434 may not have been an isolated event, and ongoing monitoring may detect the presence of additional glitches, contributing further to the body of knowledge regarding glitching pulsars. If PSR~J1706$-$4434 is seen to glitch regularly, higher cadence observations may be warranted so as to better localise each glitch in time as it occurs and hopefully measure the presence of the exponential relaxation in spin period which is typically observed immediately following a glitch event.

Future work will need to focus on developing timing solutions for those pulsars which remain unsolved. This will both help to understand the individual properties of these pulsars, but will also assist in characterising the discovered population as a whole. In the case of PSR~J1653$-$45, ongoing monitoring with high cadence bursts of observations as the pulsar passes into and out of its eclipse state will be critical in helping to constrain the pulsar's orbital and timing solution. For the black widow PSR~J1745$-$23, high-cadence observations spanning several orbits within a timescale of days or weeks will be critical in obtaining an initial orbital solution, and a timing model such as the BTX model \citep[see e.g.][]{svf+16} will likely be required to handle the long-timescale orbital variabilities inherent in a black widow system. Finally, in the cases of PSRs~J1812$-$15 and J1831$-$04, the development of binary and timing solutions will require as high a cadence of follow-up observations as can be justified in order to successfully detect them during their `on' phase. While this is likely to be an easier endeavour in the case of PSR~J1812$-$15 given its lower NF, it is important that both of these pulsars eventually be solved, as along with PSR~J1653$-$45, they represent a rarer class of long-$P$ binary pulsar. Given the significant accelerations detected in both PSR~J1812$-$15 and PSR~J1831$-$04, and an inferred $P_\text{b}$ range for each pulsar on the order of days to weeks (based on Equation~\ref{eqn: rorb}), it is possible that both PSR~J1812$-$15 and PSR~J1831$-$04 may represent objects of some scientific interest at the very least in terms of their evolutionary history.

\section*{Acknowledgements}

The Parkes Observatory is part of the Australia Telescope National Facility which is funded by the Australian Government for operation as a National Facility managed by the Commonwealth Scientific and Industrial Research Organisation (CSIRO). Pulsar research at the Jodrell Bank Centre for Astrophysics and the observations using the Lovell Radio Telescope are supported by a consolidated grant from the Science and Technology Facilities Council in the UK.  This work was supported by the ARC Centres of Excellence CE110001020 (CAASTRO) and CEA0100004 (OzGrav). Survey processing resources were provided by the Australian National Computational Infrastructure (NCI) high-performance computing centre at the Australian National University (ANU) in association with CAASTRO, and by the Max Planck Computing and Data Facility (MPCDF). The authors wish to thank Paulo Freire and Alessandro Ridolfi for their instruction and advice regarding techniques for pulsar solving and timing, Thomas Tauris for his discussions and insight into binary pulsar evolution, Eleni Graikou for her instruction on data calibration techniques, Ralph Eatough for his support and advice regarding the operation of the segmented search pipeline, and Hasan Aslan and Malte Klasing for their assistance in pulsar candidate classification. The authors further wish to acknowledge the assistance of Kira K\"{u}hn, Andreas Schmidt and the staff of the MPCDF, and Sally Cooper and the staff of the Jodrell Bank Center for Astrophysics in creation of a new tape archive copy of the HTRU-S LowLat survey at the MPCDF, which was vital in allowing for the processing and results presented in this paper. ADC acknowledges the support of both the International Max Planck Research School (IMPRS) for Astronomy and Astrophysics at the Universities of Bonn and Cologne, and the Bonn-Cologne Graduate School of Physics and Astronomy (BCGS). We further acknowledge that these results are based upon work initially presented in \cite{cameron18}. However, this paper represents a significant update, and should be seen as superseding this earlier doctoral work.




\bibliographystyle{mnras}
\bibliography{htru_xiv} 




\appendix

\section{Previously known pulsars within the survey region}

\subsection{Redetections}\label{ap-subsec: redetections}

As described in Section~\ref{sec: redetections}, \PSRindividualREDETECTIONS individual redetections of \PSRuniqueREDETECTIONS were made as part of the survey processing reported in this paper. We list the details of all \PSRindividualREDETECTIONS redetections in Table~\ref{ap-tab: redetection sample} (full version available as Supporting Information with the online version of the paper).

\begin{table*}
\begin{center}
\caption{\PSRindividualREDETECTIONS redetections of \PSRuniqueREDETECTIONS pulsars recorded during processing of the 44\,\% of the HTRU-S LowLat survey as reported in this paper. Listed for each redetection are the beam in which the detection was made (identified by a UTC time stamp and beam number), the Galactic longitude ($l$) and latitude ($b$) of the pulsar, the offset between this position and the center of the telescope beam ($\theta$), the observed spin period ($P_\text{obs}$) and DM ($\text{DM}_\text{obs}$) of the pulsar, the expected apparent flux density of the pulsar given by to Equation~\ref{eqn: flux position offset} ($S_\text{exp}$), the expected S/N of the pulsar as given by Equation~\ref{eqn: radiometer equation} ($\text{S/N}_\text{exp}$), and finally the observed S/N of the pulsar ($\text{S/N}_\text{obs}$). The following is a sample of the full table, which is available as Supporting Information with the online version of the paper.}\label{ap-tab: redetection sample}
\begin{tabular}{llrrllllrr}
\hline
PSR name & Pointing/Beam & $l$ & $b$ & $\theta$ & $P_\text{obs}$ & $\text{DM}_\text{obs}$ & $S_\text{exp}$ & $\text{S/N}_\text{exp}$ & $\text{S/N}_\text{obs}$ \\
 & & ($^\circ$) & ($^\circ$) & ($^\circ$) & (ms) & ($\text{cm}^{-3}\,\text{pc}$) & (mJy) & & \\
\hline
B1011$-$58 & 2012-09-05-20:01:28/04 & 283.706 & $-$2.144 & 0.17 & 819.924 & 379.0 & 0.361 & 108.0 & 83.9 \\
B1015$-$56 & 2012-07-23-23:59:36/09 & 282.732 & 0.341 & 0.11 & 503.462 & 433.2 & 1.540 & 354.0 & 265.3 \\
B1030$-$58 & 2012-09-24-21:37:39/04 & 285.907 & $-$0.980 & 0.19 & 464.210 & 415.2 & 0.151 & 38.6 & 40.2 \\
B1030$-$58 & 2012-09-23-22:09:55/05 & 285.907 & $-$0.980 & 0.24 & 464.210 & 416.6 & 0.052 & 13.4 & 17.5 \\
B1044$-$57 & 2012-08-05-00:16:37/10 & 287.065 & 0.733 & 0.20 & 369.427 & 239.4 & 0.143 & 25.3 & 26.9 \\
B1044$-$57 & 2012-09-05-21:14:20/08 & 287.065 & 0.733 & 0.12 & 369.428 & 240.0 & 0.527 & 92.8 & 89.7 \\
B1046$-$58 & 2012-08-02-06:52:42/07 & 287.425 & 0.577 & 0.28 & 123.714 & 128.3 & 0.103 & 15.7 & 38.4 \\
B1046$-$58 & 2012-08-05-00:16:37/03 & 287.425 & 0.577 & 0.12 & 123.714 & 127.8 & 3.170 & 484.0 & 498.7 \\
B1046$-$58 & 2012-08-05-00:16:37/10 & 287.425 & 0.577 & 0.38 & 123.714 & 129.3 & 0.004 & 0.5 & 41.9 \\
B1046$-$58 & 2012-09-24-03:00:15/10 & 287.425 & 0.577 & 0.19 & 123.714 & 128.3 & 1.080 & 139.0 & 231.7 \\
\hline
\end{tabular}
\end{center}
\end{table*}

\subsection{Non-detections}\label{ap-subsec: non detections}

As per Section~\ref{subsec: non detections}, 21 non-detections spanning 21 unique known pulsars were encountered by the partially-coherent segmented acceleration search pipeline. These pulsar non-detections are listed in Table~\ref{ap-tab: non detections}, along with the beam, angular offset ($\theta$), the expected S/N ($\text{S/N}_\text{exp}$) and the \textsc{psrcat} spin period $P_\text{cat}$ and $\text{DM}_\text{cat}$. For each non-detection, a manual fold of the relevant beam using the current \textsc{psrcat} ephemeris of the pulsar was also conducted. In 18 cases this manual analysis resulted in a redetection of the pulsar, with the resulting $\text{S/N}_\text{eph}$ also listed in Table~\ref{ap-tab: non detections}. While 9 of these non-detections have yet to be accounted for, the remaining 12 can be explained as follows:
\begin{itemize}
 \item PSRs~J1322$-$6329, J1501$-$5637, J1709$-$4342 and J1733$-$2837 all suffered from the influence of significant RFI in their corresponding observations, such that additional cleaning of the data beyond the standard RFI mitigation employed by our pipeline was required to redetect the pulsar.
 \item PSR~J1550$-$5418, a magnetar also known as AXP~1E~1547.0$-$5408, is known to experience significant radio variability including spans of apparent non-emission \citep{crh+07}, accounting for its non-detection both by the pipeline and the later ephemeris fold.
 \item PSR~J1709$-$4401 has been previously described by \cite{tjb+13} as an `intermittent' pulsar. Were it observed during an `off' state, this would account for its non-detection both by the pipeline and the later ephemeris fold.
 \item PSRs~J1524$-$5819 and J1611$-$4811 were both redetected by the ephemeris fold, but with $\text{S/N}_\text{eph} < \text{S/N}_\text{min}$. Therefore, the non-detection by the search pipeline of these two pulsars is not unexpected. Similarly, the ephemeris fold of PSR~J1509$-$5850 only resulted in $\text{S/N}_\text{eph}=9.4$ which, while technically above the $\text{S/N}_\text{min}$, is sufficiently close that it can likely be grouped into this category. 
\item PSR~J1747$-$2958 was discovered by \cite{cmgl02} in a search of the Mouse Nebula. Its interaction with this nebula, combined with a relatively-low DM, makes it susceptible to scintillation. Apparent flux variability due to this effect has already been observed in this pulsar's discovery observations. PSR~J1747$-$2958 was also not detected by \cite{ncb15}, and remains undetectable here even after the ephemeris fold.
\item With a relatively-long catalogue period of $P_\text{cat}= 1388\,\text{ms}$ and a $\text{S/N}_\text{eph}$ significantly weaker than its $\text{S/N}_\text{exp}$, the non-detection of PSR~J1819$-$1131 is attributed to likely confusion with red noise, reducing its detectability in the Fourier domain.
\item The profile of PSR~J1822$-$1617 displays a wide pulse shape with a duty cycle of $\delta\simeq14\,\%$ and evidence of a scattering tail, consistent with its high DM. This wide profile, in combination with a weak ephemeris fold of $\text{S/N}_\text{min}=10.2$, may have contributed the pulsar's non-detection.
\end{itemize}

\begin{table*}
\begin{center}
 \caption{Pulsars with $\text{S/N}_\text{exp} > \text{S/N}_\text{min} = 9$ which were not detected in the $\sim44\,\%$ of the HTRU-S LowLat survey processed by the partially-coherent segmented acceleration search pipeline for this paper. Listed is the telescope beam of each non-detection (identified by a UTC time stamp and beam number), along with the offset ($\theta$) of the pulsar from the center of the beam, the \textsc{psrcat} values of spin period ($P_\text{cat}$) and dispersion measure ($\text{DM}_\text{cat}$), the expected S/N of the pulsar as estimated from the radiometer equation as described in Section~\ref{sec: redetections} ($\text{S/N}_\text{exp}$), and the S/N resulting from a fold of beam using the \textsc{psrcat} ephemeris of the pulsar ($\text{S/N}_\text{eph}$).}\label{ap-tab: non detections}
\begin{tabular}{lllllllp{4.5cm}}
\hline
PSR name & Pointing/Beam & $\theta$ & $P_\text{cat}$ & $\text{DM}_\text{cat}$ & $\text{S/N}_\text{exp}$ & $\text{S/N}_\text{eph}$ & Comments \\
 & & ($^\circ$) & (ms) & ($\text{cm}^{-3}\,\text{pc}$) & & & \\
\hline
J1031$-$6117 & 2012-07-18-05:13:39/07 & 0.12 & 306.411 & 506.8 & 10.5 & 12.7 & Weak, close to beam FWHM, but unambiguous in ephemeris fold.\\ 
J1233$-$6344 & 2012-04-01-10:19:04/02 & 0.094 & 756.892 & 495.0 & 11.7 & 12.4 & Weak, but unambiguous in ephemeris fold. \\
J1301$-$6310 & 2011-12-22-15:34:56/09 & 0.075 & 663.830 & 86.1 & 17.8 & 15.1 & XRS \citep{pb15}. Unambiguous in ephemeris fold. \\
J1309$-$6526 & 2012-04-14-10:01:12/03 & 0.091 & 398.292 & 340.0 & 20.3 & 10.8 & Weak detection. \\
J1322$-$6329 & 2012-04-05-14:47:06/10 & 0.096 & 2764.209 & 659 & 26.6 & 13.0 & Significant RFI contamination. Pulsar only detectable in ephemeris fold after additional cleaning. \\ 
J1501$-$5637 & 2011-12-13-18:40:47/11 & 0.11 & 782.949 & 258.0 & 18.2 & 11.0 & Significant RFI contamination. \\
J1509$-$5850 & 2012-04-03-19:16:00/03 & 0.094 & 88.922 & 140.6 & 13.4 & 9.4 & HE pulsar \citep{waa+10}. Weakly detected in ephemeris fold, close to $\text{S/N}_\text{min}$. \\ 
J1524$-$5819 & 2012-08-01-10:10:14/07 & 0.096 & 961.043 & 406.6 & 15.9 & 8.8 & Weakly detected in ephemeris fold, below $\text{S/N}_\text{min}$. \\
J1550$-$5418 & 2011-07-14-05:40:23/01 & 0.022 & 2069.833 & 830.0 & 114.0$^\text{a}$ & - & Magnetar, AXP 1E 1547.0$-$5408 \citep{crh+07}. \\
J1611$-$4811 & 2011-12-27-19:03:07/08 & 0.047 & 1296.850 & 221.0 & 12.4 & 8.6 & Weakly detected in ephemeris fold, below $\text{S/N}_\text{min}$. \\
J1633$-$4805 & 2011-12-30-19:34:06/13 & 0.096 & 710.830 & 1120.0 & 15.0 & 14.4 & Unambiguous in ephemeris fold. \\
J1709$-$4342 & 2011-12-12-05:16:53/02 & 0.10 & 1735.898 & 281.0 & 15.7 & 17.0 & RFI contamination. \\
J1709$-$4401 & 2012-09-24-06:43:27/01 & 0.026 & 865.235 & 225.8 & 233.0 & - & Described as `intermittent' \citep{tjb+13}.\\
J1733$-$2837 & 2011-12-07-03:42:23/03 & 0.070 & 768.185 & 225.0 & 14.4 & 15.5 & Significant RFI contamination.\\
J1738$-$3107 & 2011-12-22-22:10:51/03 & 0.10 & 549.498 & 735.0 & 19.3 & 12.3 & Weak, but unambiguous in ephemeris fold. \\
J1747$-$2958 & 2012-08-03-09:31:42/08 & 0.061 & 98.814 & 101.5 & 21.0 & - & Discovered by \cite{cmgl02} with $S_{1400}=0.25\,\text{mJy}$. Associated with the Mouse Nebula, with flux variability likely due to interstellar scintillation. Also undetected by \cite{ncb15}. \\
J1755$-$2534 & 2012-07-26-12:48:29/02 & 0.12 & 233.541 & 590.0 & 10.6 & 11.6 & Weak, close to beam FWHM, but unambiguous in ephemeris fold. \\ 
J1801$-$2115 & 2012-03-30-22:36:23/01 & 0.078 & 438.113 & 778.8 & 14.3 & 10.2 & Weak detection. \\
J1819$-$1131 & 2012-08-05-08:17:44/02 & 0.069 & 1388.137 & 578.0 & 17.7 & 11.4 & Weak, but unambiguous in ephemeris fold. Possible red noise confusion. \\
J1819$-$1717 & 2012-07-26-14:01:59/03 & 0.12 & 393.522 & 405.0 & 14.9 & 14.6 & Unambiguous in ephemeris fold. \\
J1822$-$1617 & 2012-07-25-14:25:48/12 & 0.064 & 831.156 & 647.0 & 11.3 & 10.2 & Wide pulse profile with $\delta\simeq14\,\%$, appears to be scattered. \\
\hline
\multicolumn{8}{l}{$^\text{a}$ For pulsars with no published pulse width, an effective pulse width of $W_\text{eff}= P_\text{cat}/2$ is used to calculate $\text{S/N}_\text{exp}$.} \\
\end{tabular}
\end{center}
\end{table*}

\subsection{Binary redetections}\label{ap-subsec: binary redetections}

As per Section~\ref{subsec: binary redetections}, 17 unique binary pulsars were redetected across 28 individual survey beams. Table~\ref{ap-tab: binary redetections} lists the highest S/N detections of each pulsar in each searched segment of each beam, along with the detected acceleration ($a$), the catalogue orbital period ($P_\text{b}$) and eccentricity ($e$) of each binary as well as the calculated limiting accelerations ($a_\text{max}$ and $a_\text{min}$) along the line of sight at the epoch of each observation.   

For those pulsars detected in multiple segments of the same beam, the S/N of the detection in each subsequent segment falls by roughly $\sqrt{2}$ as the lengths of the segments are halved until the pulsar falls below the survey noise floor and is rendered undetectable. This general trend is subject to a number of minor caveats:
\begin{itemize}
 \item Due to the fact that, as listed in Table~\ref{tab: segment parameters}, the quarter-length segments were often only searched at $500 > \left|a\right| > 200\,\text{m\,s}^{-2}$ (much higher than the $a_\text{max}$ or $a_\text{min}$ of the pulsars listed in Table~\ref{ap-tab: binary redetections}), the detections in the quarter-length segments typically break this trend with significantly reduced S/N values.
 \item The full-length detections of PSRs~J1216$-$6410 and J1748$-$3009 were made at harmonics, reducing their apparent S/N.
 \end{itemize}
 
In addition, while the majority of the accelerations reported in Table~\ref{ap-tab: binary redetections} are consistent with the calculated acceleration limits of $a_\text{max}$ and $a_\text{min}$, the increasingly large ambiguity of $\dot{P}$ which occurs over shorter integration lengths (an effect exacerbated by weak pulsar detections) means that some reported acceleration values may exceed these limits. This effect is most prominent in the eighth-segment detections given their short duration of only $t_\text{int}=9\,\text{min}$.

A similar effect can be seen in PSR~J1141$-$6545 where for the two beams from each pointing in which the pulsar was detected, a discrepancy exists between the two reported half-length acceleration values. This is likely a result of the pulsar's long spin period which, as with decreasing $t_\text{int}$, increases the ambiguity of the measured $\dot{P}$ and hence also increases the ambiguity of the measured value of $a$. This effect can also be seen in the half-length detection of PSR J1822$-$0848 in beam 2012-04-02-18:07:28/13.

\begin{landscape}
\begin{table}
 \begin{center}
 \caption{Redetections of binary pulsars from $\sim44\%$ of the HTRU-S LowLat survey. Each redetected pulsar is listed along with its \textsc{psrcat} values of orbital period $P_\text{b}$ and eccentricity $e$, and the observed spin period $P_\text{obs}$ and $\text{DM}_\text{obs}$. Also listed are the calculated maximum and minimum values of acceleration ($a_\text{max}$ and $a_\text{min}$ respectively) along the line of sight at the epoch of each beam in which the pulsar was redetected (identified by a UTC time stamp and beam number). The redetections with the highest S/N from each segment group (and their corresponding acceleration $a$) are also reported.}\label{ap-tab: binary redetections}
\begin{tabular}{llllllll|ll|ll|ll|ll}
\hline
 & & & & & & & & \multicolumn{2}{c|}{Full-length} & \multicolumn{2}{c|}{Half-length} & \multicolumn{2}{c|}{Quarter-length} & \multicolumn{2}{c}{Eighth-length} \\
PSR name & $P_\text{b}$ & $e$ & $a_\text{max}$ & $a_\text{min}$ & Pointing/Beam & $P_\text{obs}$ & $\text{DM}_\text{obs}$ & S/N & $a$ & S/N & $a$ & S/N & $a$ & S/N & $a$ \\
 & (h) & & ($\text{m\,s}^{-2}$) & ($\text{m\,s}^{-2}$) & & (ms) & ($\text{cm}^{-3}\,\text{pc}$) & & ($\text{m\,s}^{-2}$) & & ($\text{m\,s}^{-2}$) & & ($\text{m\,s}^{-2}$) & & ($\text{m\,s}^{-2}$) \\
 \hline
 B1800$-$27 & 9672.7 & 0.00051 & 0.0006 & $-$0.0006 & 2011-12-11-05:59:48/03 & 334.420 & 161.3 & -$^{\text{b}}$ & - & 29.9 & $-$2.1 & 18.9 & $-$232.7 & 17.7 & $-$23.6 \\
 & & & & & 2012-10-01-06:42:41/10 & 334.412 & 165.0 & 54.8 & $-$0.2 & 41.0 & 1.0 & 16.9 & $-$211.4 & 23.6 & $-$6.6 \\
 B1820$-$11 & 8586.3 & 0.79 & 0.005 & $-$0.006 & 2011-12-23-01:58:46/13 & 279.821 & 428.0 & 91.2 & 0.6 & 65.6 & 0.03 & 28.2 & 206.4 & 34.0 & $-$23.6 \\
 & & & & & 2011-12-31-00:27:33/12 & 279.824 & 416.5 & 16.1 & $-$0.7 & 13.5 & $-$2.1 & - & - & - & -\\
 & & & & & 2012-07-22-12:33:52/09 & 279.837 & 424.9 & 14.2 & 0.3 & 12.4 & 2.1 & - & - & - & - \\
 J1822$-$0848 & 6883.9 & 0.059 & 0.002 & $-$0.002 & 2012-04-02-18:07:28/13 & 2504.472 & 189.7 & 18.6 & $-$0.2 & 13.4 & $-$17.0 & - & - & - & - \\
 & & & & & 2012-04-12-17:41:06/13 & 2504.459 & 185.5 & 67.0 & 0.06 & 47.4 & 3.1 & 27.3 & 202.1 & 25.6 & $-$40.7 \\
 J1740$-$3052 & 5544.7 & 0.58 & 0.07 & $-$0.08 & 2012-07-21-10:23:16/05 & 570.380 & 738.0 & 83.2 & $-$0.5 & 59.9 & 0.03 & 21.0 & $-$202.9 & 32.3 & $-$6.6 \\
 J1751$-$2857 & 2657.9 & 0.00013 & 0.004 & $-$0.004 & 2012-12-14-00:34:43/05 & 3.915 & 43.0 & 16.0 & 0.06 & 14.9 & 0.03 & - & - & - & - \\
 J1125$-$5825 & 1833.7 & 0.00026 & 0.009 & $-$0.009 & 2011-10-04-20:14:39/07 & 3.102 & 124.9 & 20.9 & 0.0 & 14.9 & 0.03 & - & - & - & - \\
 & & & & & 2011-12-20-18:13:43/06 & 3.102 & 124.9 & 17.5 & 0.0 & 12.7 & 0.03 & - & - & - & - \\
 & & & & & 2011-12-27-13:47:20/07 & 3.102 & 124.9 & 24.8 & 0.0 & 17.7 & 0.03 & - & - & - & - \\
 J1727$-$2946 & 967.4 & 0.046 & 0.06 & $-$0.06 & 2011-10-12-03:10:49/04 & 27.086 & 61.3 & 15.1 & 0.0 & 11.3 & 0.03 & - & - & - & - \\ 
 J1811$-$1736 & 450.7 & 0.83 & 0.98 & $-$4.8 & 2011-12-31-22:59:57/01 & 104.148 & 475.2 & 40.5 & 0.9 & 29.4 & 0.03 & 11.3 & 202.1 & 16.0 & $-$6.6 \\
 J1454$-$5846 & 298.2 & 0.0019 & 0.27 & $-$0.27 & 2012-12-30-17:34:08/06 & 45.244 & 115.6 & 19.9 & $-$0.2 & 14.6 & 0.03 & - & - & - & - \\
 J1811$-$2405 & 150.5 & 0.0000016 & 0.23 & $-$0.23 & 2012-08-05-14:33:10/04 & 2.661 & 60.8 & 24.9 & 0.3 & 22.1 & 0.03 & - & - & - & - \\
 J1337$-$6423 & 114.8 & 0.00002 & 0.91 & $-$0.91 & 2012-01-02-01:12:28/09 & 9.425 & 259.3 & 16.5 & 0.6 & - & - & - & - & - & - \\
 J1216$-$6410 & 96.9 & 0.0000068 & 0.29 & $-$0.29 & 2012-04-13-09:24:34/09 & 3.540 & 47.4 & 35.4$^{\text{a}}$ & 0.0 & 41.2 & 0.03 & - & - & 24.0 & $-$6.6 \\ 
 J1748$-$3009 & 70.4 & 0 & 0.24 & $-$0.24 & 2012-11-30-04:17:16/08 & 19.367 & 418.6 & 11.7$^{\text{a}}$ & 0.06 & - & - & - & - & - & - \\  
 J1431$-$5740 & 65.4 & 0.0000043 & 0.48 & $-$0.48 & 2011-12-23-16:52:20/07 & 4.111 & 131.3 & 18.7 & 0.3 & 12.4 & 0.03 & - & - & - & - \\
 J1435$-$6100 & 32.5 & 0.000011 & 5.3 & $-$5.3 & 2011-12-14-19:52:54/06 & 9.351 & 114.1 & - & - & 20.9 & $-$3.2 & - & - & 14.2 & $-$6.6 \\
 & & & & & 2011-12-22-16:53:19/07 & 9.350 & 114.1 & - & - & 12.5 & 3.1 & - & - & - & - \\
 J1802$-$2124 & 16.8 & 0.0000025 & 12.1 & $-$12.1 & 2011-10-12-04:24:15/07 & 12.649 & 149.9 & - & - & 16.6 & 11.6 & - & - & - & - \\
 & & & & & 2011-12-30-23:14:07/02 & 12.652 & 148.9 & 12.9 & 1.1 & 17.0 & 3.1 & - & - & - & - \\
 J1141$-$6545 & 4.7 & 0.17 & 56.5 & $-$108.9 & 2012-02-18-20:27:49/05 & 394.152 & 114.8 & 191.5 & $-$1.0 & 285.9 & $-$7.4 & 123.1 & $-$202.9 & 162.7 & $-$40.7 \\
 & & & & & 2012-02-18-20:27:49/11 & 394.029 & 118.2 & - & - & 12.7 & $-$80.8 & - & - & - & - \\
 & & & 57.5 & $-$108.2 & 2012-07-23-00:10:48/07 & 394.034 & 114.8 & 31.0 & $-$0.7 & 46.7 & $-$77.6 & 23.7 & $-$198.7 & 28.2 & $-$6.6 \\
 & & & & & 2012-07-23-00:10:48/08 & 394.151 & 117.1 & 13.3 & 0.6 & 18.3 & $-$9.6 & 8.7 & $-$202.9 & 10.8 & $-$6.6 \\
 \hline
  \multicolumn{16}{l}{$^{\text{a}}$ Indicates that the highest S/N detection for that segment was found at a harmonic.} \\
  \multicolumn{16}{l}{$^{\text{b}}$ The full-length segment for beam 2011-12-11-05:59:48/03 suffered a processing error, resulting in the non-detection of B1800$-$27 in this segment.} \\
\end{tabular}
\end{center}
\end{table}
\end{landscape}


\bsp	
\label{lastpage}
\end{document}